\documentclass[final]{IEEEtran}

\usepackage[encapsulated]{CJK}
\usepackage{ucs}
\usepackage[utf8x]{inputenc}
\usepackage[cmex10]{amsmath}
\usepackage{amsmath,amssymb,amscd,bbm,amsthm,mathrsfs,dsfont}
\usepackage{algorithmic,algorithm}
\usepackage{mdwmath}
\usepackage{mdwtab}
\usepackage{bm,upgreek}
\usepackage{cite}
\usepackage{rotating,graphics,psfrag,epsfig}
\usepackage{array}
\usepackage{booktabs}
\usepackage{indentfirst}
\usepackage{subfigure}
\usepackage{lipsum,fancyhdr,lastpage,refcount}
\usepackage[hyphens]{url}
\usepackage{fixltx2e,dblfloatfix}
\usepackage{blindtext}
\usepackage[T1]{fontenc}
\usepackage{color}

\IEEEoverridecommandlockouts

\let\oldnl\nl
\newcommand{\nonl}{\renewcommand{\nl}{\let\nl\oldnl}}

\begin{document}

\title{Multi-Tenant Cross-Slice Resource Orchestration: A Deep Reinforcement Learning Approach}

\author{\IEEEauthorblockN{Xianfu Chen, \emph{Member}, \emph{IEEE}, Zhifeng Zhao, \emph{Member}, \emph{IEEE}, Celimuge Wu, \emph{Senior Member}, \emph{IEEE}, Mehdi Bennis, \emph{Senior Member}, \emph{IEEE}, Hang Liu, \emph{Senior Member}, \emph{IEEE}, Yusheng Ji, \emph{Senior Member}, \emph{IEEE}, and Honggang Zhang, \emph{Senior Member}, \emph{IEEE}}

\thanks{X. Chen is with the VTT Technical Research Centre of Finland, Oulu, Finland (e-mail: xianfu.chen@vtt.fi).}

\thanks{Z. Zhao is with the Zhejiang Lab, and also with the College of Information Science and Electronic Engineering, Zhejiang University, Hangzhou, China (e-mail: zhaozf@\{zhejianglab.com, zju.edu.cn\}).}

\thanks{C. Wu is with the Graduate School of Informatics and Engineering, University of Electro-Communications, Tokyo, Japan (email: clmg@is.uec.ac.jp).}

\thanks{M. Bennis is with the Centre for Wireless Communications, University of Oulu, Finland (email: mehdi.bennis@oulu.fi).}

\thanks{H. Liu is with the Department of Electrical Engineering and Computer Science, the Catholic University of America, USA (e-mail: liuh@cua.edu).}

\thanks{Y. Ji is with the Information Systems Architecture Research Division, National Institute of Informatics, Tokyo, Japan (e-mail: kei@nii.ac.jp).}

\thanks{H. Zhang is with the College of Information Science and Electronic Engineering, Zhejiang University, Hangzhou, China (e-mail: honggangzhang@zju.edu.cn).}

\thanks{\emph{Corresponding author: Celimuge Wu.}}
}

\maketitle

\begin{abstract}

With the cellular networks becoming increasingly agile, a major challenge lies in how to support diverse services for mobile users (MUs) over a common physical network infrastructure.
Network slicing is a promising solution to tailor the network to match such service requests.
This paper considers a system with radio access network (RAN)-only slicing, where the physical infrastructure is split into slices providing computation and communication functionalities.
A limited number of channels are auctioned across scheduling slots to MUs of multiple service providers (SPs) (i.e., the tenants).
Each SP behaves selfishly to maximize the expected long-term payoff from the competition with other SPs for the orchestration of channels, which provides its MUs with the opportunities to access the computation and communication slices.
This problem is modelled as a stochastic game, in which the decision makings of a SP depend on the global network dynamics as well as the joint control policy of all SPs.
To approximate the Nash equilibrium solutions, we first construct an abstract stochastic game with the local conjectures of channel auction among the SPs.
We then linearly decompose the per-SP Markov decision process to simplify the decision makings at a SP and derive an online scheme based on deep reinforcement learning to approach the optimal abstract control policies.
Numerical experiments show significant performance gains from our scheme.

\end{abstract}


\section{Introduction}
\label{intr}

With the proliferation of smart mobile devices, a multitude of emerging broadband applications are driving up the demands of wireless mobile services \cite{Cisc18}.
To keep up with the demands, new cellular network infrastructures, such as small cells, are being constantly deployed \cite{Andr12}.
An advanced dense network infrastructure can reach much higher network capacity due to a shorter transmission range and a smaller number of mobile users (MUs) per cell site.
However, the coordinated control plane decisions in a dense radio access network (RAN) make it expensive to deploy and extremely complex to manage.
Meanwhile, the computation-intensive applications, e.g., augmented reality and online gaming, are gaining increasing popularity in recent years \cite{Abba18, Mao17}.
In general, the MU-end terminal devices are constrained by battery capacity and processing speed of the central processing unit (CPU).
The tension between computation-intensive applications and resource-constrained terminal devices calls for a revolution in computing infrastructure \cite{Saty17}.
Mobile-edge computing (MEC), which brings the computing capabilities within the RANs in close proximity to MUs, is envisioned as a promising solution \cite{Mao17, Wu19, Zhen19}.
Offloading a computation task to a MEC server for execution involves wireless transmissions.
Hence how to orchestrate wireless radio resources between MEC and traditional mobile services requires a careful design and adds another dimension of complexity to the network management \cite{Zhao18, Zhou17}.
By abstracting all physical base stations (BSs) in a geographical area as a logical big BS, the software-defined networking (SDN) concept provides infrastructure flexibility as well as service-oriented customization \cite{Xia15, Afol18}, and hence simplifies the management of a dense RAN \cite{Gudi13, Chen14, Chen15_1}.
In a software-defined RAN, the SDN-orchestrator handles all control plane decisions.

Another key benefit from software-defined RAN is to facilitate network sharing \cite{Lian15}.
Network sharing has been studied by 3rd Generation Partnership Project (3GPP) Technical Specification Group (TSG) Service and System Aspects (SA) 1 - \emph{Services} in \cite{3gpp14}.
Based on such a study, the sharing paradigm introduced by 3GPP TSG SA 5 - \emph{Telecom Management} considers that an infrastructure provider (InP), which is referred to as a master operator, is responsible for the configuration of a shared physical network \cite{3gpp18}.
As such, the same network infrastructure is able to host, on a RAN-as-a-service basis, multiple service providers (SPs), which is also known as multiple tenants \cite{Samd16}.
This breaks the traditional business model regarding the single ownership of a network infrastructure \cite{Fris08} and creates new business scenarios \cite{3gpp14}.
For example, an over-the-top (OTT) application provider (e.g., Netflix and Google \cite{goog18}) can become a SP so as to lease wireless radio resources from the InP to improve the Quality-of-Service (QoS) and the Quality-of-Experience (QoE) for its subscribed MUs.
In addition, a SP can also be a cellular network operator based on a long-term contractual agreement with the InP \cite{3gpp18}.
Building upon the 3GPP TSG SA 5 network sharing paradigm \cite{3gpp18}, a software-defined RAN architecture and its integration with network function virtualization enable RAN-only slicing that splits the physical RAN infrastructure into multiple virtual slices \cite{Ordo17, Sall17}.
The RAN slices can be customized for diverse service requests with various QoS and QoE requirements.
Under the context, we are concerned in this paper with a software-defined RAN where the RAN slices are specifically tailored to accommodate both computation and communication functionalities \cite{Shah17}.

From an economic viewpoint, the relationship among InP, SPs and MUs under the service-oriented RAN-only slicing fits well the business-to-business-to-consumer business model \cite{Zhou16}.
In this model, the InP takes the role as a wholesale provider, which provides the brokers with the wireless connections to the RAN slices.
The SPs (i.e., the brokers) play as the middlemen between the InP and the MUs (i.e., the consumers).
From a technical point of view, the challenges yet remain for the deployment of RAN-only slicing.
Particularly, the mechanisms that efficiently exploit the decoupling of control plane and data plane under a software-defined architecture must be developed to achieve optimized radio resource utilization across logically independent RAN slices.
For the considered software-defined RAN in this paper, the SDN-orchestrator manages a limited number of channels.
Multiple SPs compete to orchestrate channel access opportunities for their subscribed MUs, which request emerging MEC and traditional mobile services in accordance with the network dynamics.
Network dynamics originate from the mobilities as well as the random computation task and data packet arrivals of MUs.
Upon receiving the auction bids from all SPs, the SDN-orchestrator allocates channels to MUs through a Vickrey-Clarke-Groves (VCG) pricing mechanism\footnote{One major advantage of the VCG mechanism is that the dominant auction policy for a SP is to bid the true values for the channels.} \cite{Ji07}.
Each MU then proceeds to offload computation tasks and schedule data packets over the assigned channel with the objective of optimizing the expected long-term performance.
%
%
The main technical contributions from this paper are listed.
\begin{itemize}
  \item We formulate the multi-tenant cross-slice radio resource orchestration problem as a non-cooperative stochastic game under the multi-agent Markov decision process (MDP), in which each SP interacts with other competing SPs in the network and aims to selfishly maximize its own expected long-term payoff.
  \item Without any information exchange among the SPs, we transform the stochastic game into an abstract stochastic game with a bounded performance regret.
  \item We further put forward a linear decomposition approach to solve the per-SP MDP, leading to simplified decision makings.
      The decomposition approach allows a MU of each SP to locally compute the state-value functions.
  \item To deal with the huge local state space faced by a MU, we leverage a deep reinforcement learning (DRL) algorithm \cite{Hass16} to learn the optimal computation offloading and packet scheduling policies without a priori statistical knowledge of the network dynamics.
  \item Numerical experiments using TensorFlow \cite{Abad16} are carried out to verify the theoretical studies in this paper, showing that our proposed scheme outperforms three state-of-the-art baseline schemes.
\end{itemize}

The remainder of this paper is organized as follows.
In the next section, we briefly review the related works in the literature.
In Section \ref{sysm}, we describe the considered system model and the assumptions used throughout this paper.
In Section \ref{prob}, we formulate the problem of competitive multi-tenant cross-slice resource orchestration as a stochastic game and discuss the general best-response solution.
In Section \ref{probSolv}, we propose to approximate the stochastic game by an abstract stochastic game and derive an online learning scheme to approach the optimal decision makings.
In Section \ref{simu}, we provide numerical experiments under various settings to compare the performance from our scheme with other state-of-the-art baselines.
Finally, we draw the conclusions in Section \ref{conc}.
For convenience, Table \ref{tabl1} summarizes the major notations of this paper.
\begin{table}[t]
  \caption{Major notations used in the paper.}\label{tabl1}
        \begin{center}
        \begin{tabular}{c|l}
              \hline
              $I$/$\mathcal{I}$                                     & number/set of SPs                                                     \\\hline
              $B$/$\mathcal{B}$                                     & number/set of BSs                                                     \\\hline
              $\mathcal{L}_b$                                       & set of locations covered by BS $b$                                    \\\hline
              $|\mathcal{N}_i|$/$\mathcal{N}_i$                     & number/set of MUs subscribed to SP $i$                                \\\hline
              $\delta$                                              & time duration of one scheduling slot                                  \\\hline
              $J$/$\mathcal{J}$                                     & number/set of channels                                                \\\hline
              $\eta$                                                & bandwidth of a channel                                                \\\hline
              $\bm\beta_i$, $\bm\beta_i^k$                          & bid of SP $i$                                                         \\\hline
              $\nu_i$, $\nu_i^k$                                    & true value of SP $i$                                                  \\\hline
              $\mathbf{C}_i$, $\mathbf{C}_i^k$                      & channel request profile of SP $i$                                     \\\hline
              $C_{b, i}$, $C_{b, i}^k$                              & channels needed from SP $i$ in coverage of BS $b$                     \\\hline
              $\bm\rho_n$, $\bm\rho_n^k$                            & channel allocation vector of MU $n$                                   \\\hline
              $\rho_{n, j}$, $\rho_{n, j}^k$                        & channel allocation indicator of MU $n$ over channel $j$               \\\hline
              $\varphi_n$, $\varphi_n^k$                            & channel allocation variable of MU $n$                                 \\\hline
              $\bm\phi$, $\bm\phi^k$                                & auction winner determination vector                                   \\\hline
              $\phi_i$, $\phi_i^k$                                  & auction winner indicator of SP $i$                                    \\\hline
              $\tau_i$, $\tau_i^k$                                  & payment of SP $i$                                                     \\\hline
              $L_n$, $L_n^k$                                        & location of MU $n$                                                    \\\hline
              $H_n$, $H_n^k$                                        & channel gain of MU $n$                                                \\\hline
              $A_{n, (\mathrm{t})}$, $A_{n, (\mathrm{t})}^k$        & task arrivals of MU $n$                                               \\\hline
              $\mu_{(\mathrm{t})}$                                  & input data size of a computation task                                 \\\hline
              $\vartheta$                                           & required CPU cycles per input bit                                     \\\hline
              $R_{n, (\mathrm{t})}$, $R_{n, (\mathrm{t})}^k$        & computation offloading decision                                       \\\hline
              $\mu_{(\mathrm{p})}$                                  & data packet size                                                      \\\hline
              $W_n$, $W_n^k$                                        & queue state of MU $n$                                                 \\\hline
              $W^{(\max)}$                                          & maximum queue length                                                  \\\hline
              $A_{n, (\mathrm{p})}$, $A_{n, (\mathrm{p})}^k$        & packet arrivals of MU $n$                                             \\\hline
              $R_{n, (\mathrm{p})}$, $R_{n, (\mathrm{p})}^k$        & packet scheduling decision of MU $n$                                  \\\hline
              $P_{n, (\mathrm{tr})}^k$                              & transmit energy consumption of MU $n$ at slot $k$                     \\\hline
              $\Omega^{(\max)}$                                     & maximum transmit power of a MU                                        \\\hline
              $P_{n, (\mathrm{CPU})}^k$                             & CPU energy consumption of MU $n$ at slot $k$                          \\\hline
              $\varrho$                                             & CPU-cycle frequency of a MU                                           \\\hline
              $\bm\chi_n$, $\bm\chi_n^k$                            & local network state of MU $n$                                         \\\hline
              $\bm\pi_i$, $\tilde{\bm\pi}_i$                        & control policy of SP $i$                                              \\\hline
              $\pi_{i, (\mathrm{c})}$,
              $\tilde{\pi}_{i, (\mathrm{c})}$                       & channel auction policy of SP $i$                                      \\\hline
              $\bm\pi_{i, (\mathrm{t})}$                            & computation offloading policy of SP $i$                               \\\hline
              $\bm\pi_{i, (\mathrm{p})}$                            & packet scheduling policy of SP $i$                                    \\\hline
              $F_i$                                                 & payoff function of SP $i$                                             \\\hline
              $U_n$                                                 & utility function of MU $n$                                            \\\hline
              $\alpha_n$                                            & utility price for MU $n$                                              \\\hline
              $\ell_n$                                              & weight of energy consumption of MU $n$                                \\\hline
              $D_n$, $D_n^k$                                        & packet drops at MU $n$                                                \\\hline
              $V_i$                                                 & expected long-term payoff of SP $i$                                   \\\hline
              $\mathds{V}_i$                                        & optimal state-value function of SP $i$                                \\\hline
              $\mathds{U}_n$                                        & expected long-term utility of MU $n$                                  \\\hline
              $\mathds{U}_i$                                        & expected long-term payment of SP $i$                                  \\\hline
              $s_i$, $s_i^k$                                        & abstract state at SP $i$                                              \\\hline
              $Q_n$                                                 & state action-value of MU $n$                                          \\\hline
              $\bm\theta_n$, $\bm\theta_n^k$, $\bm\theta_{n, -}^k$  & parameters associated with the DQN of MU $n$                          \\\hline
              $\mathcal{M}_n^k$                                     & replay memory of MU $n$ at slot $k$                                   \\\hline
              $\mathcal{O}_n^k$                                     & mini-batch of MU $n$ at slot $k$                                      \\
              \hline
        \end{tabular}
    \end{center}
\end{table}

\section{Related Works}

Network slicing has been introduced as one of the key features of the International Mobile Telecommunication (IMT)-2020 network \cite{ITU18}.
Basically, we have three types of network slicing implementation as in previous work \cite{Zhou16}, namely:
\begin{enumerate}
  \item Core network (CN)-only slicing -- The network slicing is performed only on the CN, and neither the RAN nor the MUs need to be configured for a selected slice of the CN.
      After the MUs being attached to the correct CN slice, all the interfaces and the procedures between the RAN and the CN control planes remain unchanged.
  \item RAN-only slicing -- The RAN slices run on a wireless platform containing the radio hardware and the baseband resource pool, which exhibit less elasticity compared with the CN-only slicing.
      Different slices of a RAN applies different air interface parameters to cater for heterogeneous mobile service requests.
  \item CN-RAN slicing -- With this type of network slicing, each slice of a RAN is connected to a specific slice of the CN.
      Once a MU obtains the access to a RAN slice, it does not need to select the CN slice, which indicates that the slice selection procedure is the same as that of the RAN-only network slicing.
      This type of network slicing brings the advantages of being able to program the functionalities of the CN slices as well as to customize the air interfaces of the RAN slices.
\end{enumerate}
In this paper, we consider RAN-only slicing since the interplay among the InP, the competing SPs and the MUs exists on the RAN part of the whole network and the SDN-orchestrator makes all control plane decisions.

There exist a number of research efforts on resource allocation for network slicing.
Game theory has been useful as an analytical framework for assessing the multi-tenant resource allocation performance in network slicing \cite{Han11, Han15}.
In \cite{Dats18}, Datsika et al. introduced a matching theoretic flow prioritization algorithm that respects network neutrality for the network resource scheduling problem, in which the OTT service providers interact with the InP.
In \cite{Xiao18}, Xiao et al. formulated a network slicing game based on the overlapping coalition formation game to investigate the potential cooperation among the cellular network operators.
In \cite{Caba19}, Caballero et al. analyzed a ``share-constrained proportional allocation'' mechanism for resource sharing to realize network slicing, which falls into a network slicing game framework.
In \cite{DOro18}, D'Oro et al. designed a near-optimal low-complexity distributed algorithm to settle down the problem of RAN slicing, which was modelled as a congestion game.
In \cite{Sun19}, Sun et al. established a Stackelberg game to describe the interplay among the global radio resource manage, the local radio resource managers and the MUs in fog RAN slicing.
All above works fail to adequately characterize the long-term resource allocation performance.

To achieve long-term performance optimization, the resource allocation in network slicing should account for the network dynamics.
In \cite{Xiao1802}, Xiao et al. studied dynamic network slicing for a fog computing system under randomly fluctuating energy harvesting and workload arrival processes and proposed a Bayesian learning approach to achieve the optimal resource slicing structure among the fog nodes.
The approach in this work relies on the statistics of network dynamics.
To alleviate the requirement of a priori statistical knowledge of network dynamics, Fu and Kozat developed an online reinforcement learning algorithm to solve the optimal policy for the stochastic game which models the non-cooperative behaviours of SPs during the competition for transmission rate allocation \cite{Fu13}.
However, these efforts concentrate on the traditional mobile services and are constrained by only either infrastructure slicing or spectrum resource slicing \cite{Xiao18}.
On the other hand, network slicing assisted by intelligent learning enables adaptability and robustness to a dynamic networking environment \cite{Jian17}.
Machine learning techniques are currently being actively discussed by several standards development organizations and industrial forums, for example, International Telecommunication Union--``Focus Group on Machine Learning for Future Networks including 5G'' \cite{ML5G}, European Telecommunications Standards Institute--Industrial Specification Group ``Experiential Networked Intelligence (ENI)'' \cite{ETSI}, International Organization for Standardization--``Artificial Intelligence'' \cite{ISO} and TeleManagement Forum--Catalyst Project ``Artificial Intelligence makes Smart BPM Smarter'' \cite{TM}.

\section{System Descriptions and Assumptions}
\label{sysm}

\begin{figure}[t]
  \centering
  \includegraphics[width=21pc]{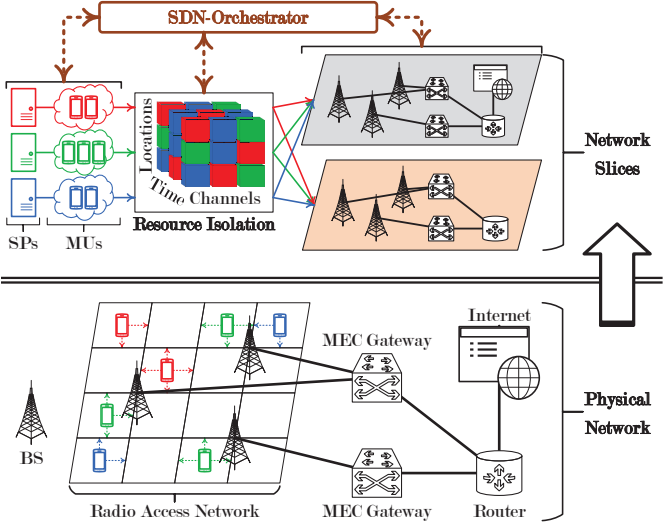}
  \caption{Architecture of the radio access network-only slicing (BS: base station; SDN: software-defined networking.).
  A physical network infrastructure managed by an infrastructure provider is split into virtual slices with functionalities particularly designed for serving computation and communication requirements.
  Multiple service providers (SPs), which is also known as the tenants, provide both emerging mobile-edge computing (MEC) and traditional mobile services.
  The mobile users (MUs) of SPs, which are shown in the different colors, move across the service region.
  Over the time horizon, the SDN-orchestrator allocates the limited wireless radio resource to MUs for the access to two network slices based on the bids that are submitted by their respective subscribing SPs.}
  \label{systMode}
\end{figure}

As being illustrated in Fig. \ref{systMode}, this paper considers a system with RAN-only slicing.
The physical network infrastructure is split into virtual computation and communication slices tailored to heterogeneous mobile service requests, which can be basically categorized as the emerging MEC and the traditional mobile services.
The shared RAN, which consists of a set $\mathcal{B}$ of physical BSs, covers a service region with a set $\mathcal{L}$ of locations (or small areas) with each being characterized by uniform signal propagation conditions \cite{Ho14, Chen15}.
We choose $\mathcal{L}_b$ to designate the set of locations covered by a BS $b \in \mathcal{B}$.
For any two BSs in the RAN, we assume that $\mathcal{L}_b \cap \mathcal{L}_{b'} = \emptyset$, where $b' \in \mathcal{B}$ and $b' \neq b$.
We represent the geographical distribution of BSs by a topological graph $\mathcal{TG} = \langle\mathcal{B}, \mathcal{E}\rangle$, where $\mathcal{E} = \{e_{b, b'}: b \neq b', b, b' \in \mathcal{B}\}$ represents the relative locations between the BSs with
\begin{align}
     e_{b, b'}
   = \left\{\!\!
     \begin{array}{l@{~}l}
        1, & \mbox{if BSs } b \mbox{ and } b' \mbox{ are neighbours};   \\
        0, & \mbox{otherwise}.
     \end{array}
     \right.
\end{align}
Different SPs provide different mobile services, and each MU can subscribe to only one SP $i \in \mathcal{I} = \{1, \cdots, I\}$.
Let $\mathcal{N}_i$ be the set of MUs of SP $i$, then $\mathcal{N} = \cup_{i \in \mathcal{I}} \mathcal{N}_i$ denotes the set of all MUs across the whole network.

\subsection{Inter-Tenant Channel Auction}

We consider a system with a set $\mathcal{J} = \{1, \cdots, J\}$ of non-overlapping orthogonal channels with the same bandwidth $\eta$ (in Hz).
The whole system operates across discrete scheduling slots, each of which is indexed by an integer $k \in \mathds{N}_+$ and is assumed to be of equal time duration $\delta$ (in seconds).
Over the time horizon, the MUs move in the service region $\mathcal{L}$ following a Markov mobility model.
Such a mobility model is widely used in the literature \cite{Nich08, Cheu15}.
Let $\mathcal{N}_{b, i}^k$ be the set of MUs appearing in the coverage of a BS $b \in \mathcal{B}$ at a scheduling slot $k \in \mathds{N}_+$ that are subscribed to SP $i \in \mathcal{I}$, then $\mathcal{N}_i = \cup_{b \in \mathcal{B}} \mathcal{N}_{b, i}^k$, $\forall k \in \mathds{N}_+$.
We assume that during a scheduling slot, a MU at a location can only be associated with the BS that covers the location.
The SPs compete for the limited number of channels in order to provide their MUs the access to the virtual computation and communication slices.
Specifically, at the beginning of each scheduling slot $k$, each SP $i$ submits to the SDN-orchestrator a bid given by $\hat{\bm\beta}_i^k = (\hat{\nu}_i^k, \hat{\mathbf{C}}_i^k)$, which is not necessarily equal to $\bm\beta_i^k = (\nu_i^k, \mathbf{C}_i^k)$.
Herein, $\mathbf{C}_i^k = (C_{b, i}^k: b \in \mathcal{B})$ with $C_{b, i}^k$ being the number of potentially needed channels within the coverage of a BS $b$ and $\nu_i^k$ is the true value over $\mathbf{C}_i^k$.
Upon receiving the auction bids $\hat{\bm\beta}^k = (\hat{\bm\beta}_i^k: i \in \mathcal{I})$ from all SPs, the SDN-orchestrator proceeds to allocate the channels to MUs and computes the payment $\tau_i^k$ for each SP $i$.
Let $\bm\rho_n^k = (\rho_{n, j}^k: j \in \mathcal{J})$ be the channel allocation vector for a MU $n \in \mathcal{N}$, where
\begin{align}
     \rho_{n, j}^k
   = \left\{\!\!
     \begin{array}{l@{~}l}
        1, & \mbox{if channel } j \mbox{ is allocated to}                                   \\
           & \mbox{MU } n \in \mathcal{N} \mbox{ at scheduling slot } k;                    \\
        0, & \mbox{otherwise}.
     \end{array}
     \right.
\end{align}
We apply the following constraints for the centralized channel allocation at the SDN-orchestrator during a single slot,
\begin{align}
        \left(\sum_{i \in \mathcal{I}} \sum_{n \in \mathcal{N}_{b, i}^k}  \rho_{n, j}^k\right) \cdot
 & \!   \left(\sum_{i \in \mathcal{I}} \sum_{n \in \mathcal{N}_{b', i}^k} \rho_{n, j}^k\right) = 0,         \nonumber\\
        \mbox{if } e_{b, b'}
 & =    1, \forall e_{b, b'} \in \mathcal{E}, \forall j \in \mathcal{J};                                    \label{c1}\\
        \sum_{i \in \mathcal{I}} \sum_{n \in \mathcal{N}_{b, i}^k} \rho_{n, j}^k
 & \leq 1, \forall b \in \mathcal{B}, \forall j \in \mathcal{J};                                            \label{c2}\\
        \sum_{j \in \mathcal{J}} \rho_{n, j}^k
 & \leq 1, \forall b \in \mathcal{B}, \forall i \in \mathcal{I}, \forall n \in \mathcal{N}_{b, i},          \label{c3}
\end{align}
to ensure that a channel cannot be allocated to the coverage areas of two adjacent BSs in order to avoid interference in data transmissions, and in the coverage of a BS, a MU can be assigned at most one channel and a channel can be assigned to at most one MU.
As we will see later in this paper, such assumptions make the decision makings from SPs only coupled during the channel auctions.

We denote by $\bm\phi^k = (\phi_i^k: i \in \mathcal{I})$ the winner determination in the channel auction at a scheduling slot $k$, where $\phi_i^k = 1$ if SP $i$ wins the channel auction while $\phi_i^k = 0$ indicates that no channel is allocated to the MUs of SP $i$ during the slot.
The SDN-orchestrator calculates $\bm\phi^k$ through the VCG mechanism that maximizes the true value of all SPs\footnote{Other non-VCG auction mechanisms (e.g., \cite{Jia09}) can also be implemented but do not affect the proposed resource orchestration scheme in this paper.},
\begin{equation}\label{chanSche}
  \begin{array}{cl}
                  & \bm\phi^k = \underset{\bm\phi}{\arg\max} \displaystyle\sum\limits_{i \in \mathcal{I}} \phi_i \cdot \hat{\nu}_i^k    \\
    \mathrm{s.t.} & \mbox{constraints (\ref{c1}), (\ref{c2}) and (\ref{c3})};                                                           \\
                  & \displaystyle\sum_{n \in \mathcal{N}_{b, i}^k} \varphi_n^k = \phi_i \cdot C_{b, i}^k,
                    \forall b \in \mathcal{B}, \forall i \in \mathcal{I},
  \end{array}
\end{equation}
where $\bm\phi = (\phi_i \in \{0, 1\}: i \in \mathcal{I})$ and $\varphi_n^k = \sum_{j \in \mathcal{J}} \rho_{n, j}^k$ is a channel allocation variable that equals $1$ if MU $n$ is assigned a channel and $0$, otherwise.
Moreover, the payment for each SP $i$ is calculated as
\begin{align}\label{paymCalc}
   \tau_i^k = \max\limits_{\bm\phi_{-i}} \displaystyle\sum\limits_{i' \in \mathcal{I} \setminus \{i\}} \phi_{i'} \cdot \hat{\nu}_{i'}^k
            - \max\limits_{\bm\phi} \displaystyle\sum\limits_{i' \in \mathcal{I} \setminus \{i\}} \phi_{i'} \cdot \hat{\nu}_{i'}^k,
\end{align}
where $-i$ denotes all the other SPs in $\mathcal{I}$ without the presence of SP $i$.
The economic properties of the VCG-based channel auction at a scheduling slot $k$ are as follows.
\begin{itemize}
  \item \emph{Efficiency --} When all SPs announce their true bids, the SDN-orchestrator allocates the channels to maximize the sum of values, resulting in efficient channel utilization.
  \item \emph{Individual Rationality --} Each SP $i$ can expect a nonnegative payoff $\hat{\nu}_i^k - \tau_i^k$ at any scheduling slot $k$.
  \item \emph{Truthfulness --} No SP can improve its payoff by bidding differently from its true value, which implies that the optimal bid at any slot $k$ is $\hat{\bm\beta}_i^k = \bm\beta_i^k$, $\forall i \in \mathcal{I}$.
\end{itemize}

\subsection{Computation and Communication Models}

Let $L_n^k \in \mathcal{L}$ be the location of a MU $n \in \mathcal{N}$ during a scheduling slot $k$, and the average channel gain $H_n^k = h(L_n^k)$ experienced by MU $n$ during the slot is determined by the physical distance between the MU and the associated BS\footnote{It's straightforward that given the mobility model, the average overhead of a MU incurred during inter-BS handovers is fixed.} \cite{Ho14, Chen15}.
At the beginning of each scheduling slot $k$, each MU $n$ independently generates a random number $A_{n, (\mathrm{t})}^k \in \mathcal{A} = \{0, 1, \cdots, A_{(\mathrm{t})}^{(\max)}\}$ of computation tasks.
We represent a computation task by $(\mu_{(\mathrm{t})}, \vartheta)$ with $\mu_{(\mathrm{t})}$ and $\vartheta$ being, respectively, the input data size (in bits) and the number of CPU cycles required to accomplish one input bit of the computation task.
This work assumes that the task arrival sequence $\{A_{n, (\mathrm{t})}^k: k \in \mathds{N}_+\}$ follows a Markov process \cite{He17}.
Two options are available for each computation task\footnote{The kind of computation tasks that can only be processed at the mobile devices \cite{He17} does not affect the optimization goal and hence is neglected.}: 1) being processed locally at the MU; and 2) being offloaded to the logical MEC gateway in the computation slice.
In other words, the arriving computation tasks must be executed during the scheduling slot\footnote{For simplicity, we assume that the CPU power at a mobile device matches the maximum computation task arrivals and a MU can hence process $A_{(\mathrm{t})}^{(\max)}$ tasks within one scheduling slot.}.
The computation offloading decision for MU $n$ at a slot $k$ specifies the number $R_{n, (\mathrm{t})}^k$ of tasks to be transmitted to the MEC server.
The final number of tasks to be processed by the mobile device hence is $A_{n, (\mathrm{t})}^k - \varphi_n^k \cdot R_{n, (\mathrm{t})}^k$.
Meanwhile, a data queue is maintained at each MU to buffer the packets coming from the traditional mobile service.
The arriving packets get queued until transmissions and we assume that every data packet has a constant size of $\mu_{(\mathrm{p})}$ (bits).
Let $W_n^k$ and $A_{n, (\mathrm{p})}^k$ be, respectively, the queue length and the random new packet arrivals for MU $n$ at the beginning of slot $k$.
The packet arrival process is assumed to be independent among the MUs and identical and independently distributed across the scheduling slots.
Let $R_{n, (\mathrm{p})}^k$ be the number of data packets that are to be removed from the queue of MU $n$ at slot $k$.
Then the number of packets that are eventually transmitted via the communication slice is $\varphi_n^k \cdot R_{n, (\mathrm{p})}^k$, and the queue evolution of MU $n$ can be written in the form of
\begin{align}
  W_n^{k + 1} = \min\!\left\{W_n^k - \varphi_n^k \cdot R_{n, (\mathrm{p})}^k + A_{n, (\mathrm{p})}^k, W^{(\max)}\right\},
\end{align}
where $W^{(\max)}$ is the maximum buffer size that restricts $W_n^k \in \mathcal{W} = \{0, \cdots, W^{(\max)}\}$.

Following the discussions in \cite{Berr02}, the energy (in Joules) consumed by a MU $n \in \mathcal{N}$ for reliably transmitting input data of $\varphi_n^k \cdot R_{n, (\mathrm{t})}^k$ computation tasks and $\varphi_n^k \cdot R_{n, (\mathrm{p})}^k$ packets during a scheduling slot $k$ can be calculated as
\begin{align}\label{tranPowe}
   P_{n, (\mathrm{tr})}^k =
   \frac{\delta \cdot \eta \cdot \sigma^2} {H_n^k} \cdot \left(2^{\frac{\varphi_n^k \cdot \left(\mu_{(\mathrm{t})} \cdot R_{n, (\mathrm{t})}^k + \mu_{(\mathrm{p})} \cdot R_{n, (\mathrm{p})}^k\right)} {\eta \cdot \delta}} - 1\right),
\end{align}
where $\sigma^2$ is the noise power spectral density.
Let $\Omega^{(\max)}$ be the maximum transmit power for all MUs, namely, $P_{n, (\mathrm{tr})}^k \leq \Omega^{(\max)} \cdot \delta$, $\forall n$ and $\forall k$.
For the rest number $A_{n, (\mathrm{t})}^k - \varphi_n^k \cdot R_{n, (\mathrm{t})}^k$ of computation tasks that are processed at the mobile device of MU $n$, the CPU energy consumption is given by
\begin{align}\label{taskEner}
  P_{n, (\mathrm{CPU})}^k = \varsigma \cdot \mu_{(\mathrm{t})} \cdot \vartheta \cdot \varrho^2 \cdot \left(A_{n, (\mathrm{t})}^k - \varphi_n^k \cdot R_{n, (\mathrm{t})}^k\right),
\end{align}
where $\varsigma$ is the effective switched capacitance that depends on chip architecture of the mobile device \cite{Burd96} and $\varrho$ is the CPU-cycle frequency at a mobile device.

\subsection{Control Policy}

We denote $\bm\chi_n^k = (L_n^k, A_{n, (\mathrm{t})}^k, W_n^k) \in \mathcal{X} = \mathcal{L} \times \mathcal{A} \times \mathcal{W}$ as the local network state observed at a MU $n \in \mathcal{N}$.
Thus $\bm\chi^k = (\bm\chi_n^k: n \in \mathcal{N}) \in \mathcal{X}^{|\mathcal{N}|}$ characterizes the global state of the network, where $|\mathcal{N}|$ means the cardinality of the set $\mathcal{N}$.
Each SP $i \in \mathcal{I}$ aims to design a control policy $\bm\pi_i = (\pi_{i, (\mathrm{c})}, \bm\pi_{i, (\mathrm{t})}, \bm\pi_{i, (\mathrm{p})})$, where $\pi_{i, (\mathrm{c})}$, $\bm\pi_{i, (\mathrm{t})} = (\pi_{n, (\mathrm{t})}: n \in \mathcal{N}_i)$ and $\bm\pi_{i, (\mathrm{p})} = (\pi_{n, (\mathrm{p})}: n \in \mathcal{N}_i)$ are the channel auction, the computation offloading and the packet scheduling policies, respectively.
Note that the computation offloading policy $\pi_{n, (\mathrm{t})}$ as well as the packet scheduling policy $\pi_{n, (\mathrm{p})}$ are MU-specified, hence both $\bm\pi_{i, (\mathrm{t})}$ and $\bm\pi_{i, (\mathrm{p})}$ depend only on $\bm\chi_i^k = (\bm\chi_n^k: n \in \mathcal{N}_i) \in \mathcal{X}_i = \mathcal{X}^{|\mathcal{N}_i|}$.
The joint control policy of all SPs is given by $\bm\pi = (\bm\pi_i: i \in \mathcal{I})$.
With the observation of $\bm\chi^k$ at the beginning of each scheduling slot $k$, SP $i$ announces the auction bid $\bm\beta_i^k$ to the SDN-orchestrator for channel allocation and decides the numbers of computation tasks $\mathbf{R}_{i, (\mathrm{t})}^k$ to be offloaded and packets $\mathbf{R}_{i, (\mathrm{p})}^k$ to be transmitted following $\bm\pi_i$, that is, $\bm\pi_i(\bm\chi^k) = (\pi_{i, (\mathrm{c})}(\bm\chi^k), \bm\pi_{i, (\mathrm{t})}(\bm\chi_i^k), \bm\pi_{i, (\mathrm{p})}(\bm\chi_i^k)) = (\bm\beta_i^k, \mathbf{R}_{i, (\mathrm{t})}^k, \mathbf{R}_{i, (\mathrm{p})}^k)$, where $\mathbf{R}_{i, (\mathrm{t})}^k = (R_{n, (\mathrm{t})}^k: n \in \mathcal{N}_i)$ and $\mathbf{R}_{i, (\mathrm{p})}^k = (R_{n, (\mathrm{p})}^k: n \in \mathcal{N}_i)$.
We define an instantaneous payoff function for SP $i \in \mathcal{I}$ at a slot $k$ as below,
\begin{align}\label{payOff}
 & F_i\!\left(\bm\chi^k, \bm\varphi_i^k, \mathbf{R}_{i, (\mathrm{t})}^k, \mathbf{R}_{i, (\mathrm{p})}^k\right)                                      \nonumber\\
 & = \sum_{n \in \mathcal{N}_i} \alpha_n \cdot U_n\!\left(\bm\chi_n^k, \varphi_n^k, R_{n, (\mathrm{t})}^k, R_{n, (\mathrm{p})}^k\right) - \tau_i^k,
\end{align}
where $\bm\varphi_i^k = (\varphi_n^k: n \in \mathcal{N}_i)$ and $\alpha_n \in \mathds{R}_+$ can be treated herein as the unit price to charge a MU $n$ for realizing utility $U_n(\bm\chi_n^k, \varphi_n^k, R_{n, (\mathrm{t})}^k, R_{n, (\mathrm{p})}^k)$ from consuming power to process the arriving computation tasks and transmit the queued packets to avoid the packet overflows, which is chosen to be
\begin{align}\label{util}
  &   U_n\!\left(\bm\chi_n^k, \varphi_n^k, R_{n, (\mathrm{t})}^k, R_{n, (\mathrm{p})}^k\right)
    = U_n^{(1)}\!\left(W_n^{k + 1}\right) + U_n^{(2)}\!\left(D_n^k\right) +                                                           \nonumber\\
  &   \ell_n \cdot \left(U_n^{(3)}\!\left(P_{n, (\mathrm{CPU})}^k\right) + U_n^{(4)}\!\left(P_{n, (\mathrm{tr})}^k\right)\right),
\end{align}
where $D_n^k = \max\{W_n^k - \varphi_n^k \cdot R_{n, (\mathrm{p})}^k + A_{n, (\mathrm{p})}^k - W^{(\max)}, 0\}$ defines the number of packet drops that occur when the queue vacancy is less than the number of arriving packets,
the positive and monotonically decreasing functions $U_n^{(1)}(\cdot)$, $U_n^{(2)}(\cdot)$, $U_n^{(3)}(\cdot)$ and $U_n^{(4)}(\cdot)$ measure the satisfactions of the packet queuing delay, the packet drops, the CPU energy consumption and the transmit energy consumption,
and $\ell_n \in \mathds{R}_+$ is a constant weighting factor that balances the importance of the energy consumption within a scheduling slot.

\section{Problem Statement and Game-Theoretic Solution}
\label{prob}

In this section, we first formulate the problem of cross-slice resource orchestration among the non-cooperative SPs (i.e., the competitive channel auction, computation offloading and packet scheduling) across the time horizon as a stochastic game and then discuss the best-response solution from a game-theoretic perspective.

\subsection{Stochastic Game Formulation}

Due to the limited radio resource and the stochastic nature in networking environment, we therefore formulate the problem of cross-slice resource orchestration among multiple non-cooperative SPs over the time horizon as a stochastic game, $\mathcal{SG}$, in which $I$ SPs are the competitive players and there are a set $\mathcal{X}^{|\mathcal{N}|}$ of global network states and a collection of control policies $\{\bm\pi_i: \forall i \in \mathcal{I}\}$.
The joint control policy $\bm\pi$ induces a probability distribution over the sequence of global network states $\{\bm\chi^k: k \in \mathds{N}_+\}$ and the sequences of per-slot instantaneous payoffs $\{F_i(\bm\chi^k, \bm\varphi_i^k, \mathbf{R}_{i, (\mathrm{t})}^k, \mathbf{R}_{i, (\mathrm{p})}^k): k \in \mathds{N}_+\}$, $\forall i \in \mathcal{I}$.
From assumptions on the mobility of a MU and the random computation task and data packet arrivals, the randomness lying in $\{\bm\chi^k: k \in \mathds{N}_+\}$ is hence Markovian with the following state transition probability
\begin{align}\label{statTranProb}
   & \mathbb{P}\!\left(\bm\chi^{k + 1} | \bm\chi^k, \bm\varphi\!\left(\bm\pi_{(\mathrm{c})}\!\left(\bm\chi^k\right)\right),
        \bm\pi_{(\mathrm{t})}\!\left(\bm\chi^k\right), \bm\pi_{(\mathrm{p})}\!\left(\bm\chi^k\right)\right)                                 \\
 = & \prod_{n \in \mathcal{N}} \mathbb{P}\!\left(L_n^{k + 1} | L_n^k\right) \cdot
        \mathbb{P}\!\left(A_{n, (\mathrm{t})}^{k + 1} | A_{n, (\mathrm{t})}^k\right) \cdot                                                  \nonumber\\
   & \mathbb{P}\!\left(W_n^{k + 1} | W_n^k, \varphi_n\!\left(\bm\pi_{(\mathrm{c})}\!\left(\bm\chi^k\right)\right),
        \pi_{n, (\mathrm{t})}\!\left(\bm\chi_n^k\right), \pi_{n, (\mathrm{p})}\!\left(\bm\chi_n^k\right)\right),                            \nonumber
\end{align}
where $\mathbb{P}(\cdot)$ denotes the probability of an event, $\bm\varphi = (\bm\varphi_i: i \in \mathcal{I})$ is the global channel allocation by the SDN-orchestrator, while $\bm\pi_{(\mathrm{c})} = (\pi_{i, (\mathrm{c})}: i \in \mathcal{I})$, $\bm\pi_{(\mathrm{t})} = (\bm\pi_{i, (\mathrm{t})}: i \in \mathcal{I})$ and $\bm\pi_{(\mathrm{p})} = (\bm\pi_{i, (\mathrm{p})}: i \in \mathcal{I})$ are, respectively, the joint channel auction, the joint computation offloading and the joint packet scheduling policies.

Taking expectation with respect to the sequence of per-slot instantaneous payoffs, the expected long-term payoff\footnote{Due to the non-cooperative behaviours among SPs, the control policies, $\bm\pi_i$, $\forall i \in \mathcal{I}$, are not unichain. Therefore, our considered Markovian system is non-ergodic, for which we define an expected infinite-horizon discounted payoff function as the optimization goal of a SP.} of a SP $i \in \mathcal{I}$ for a given initial global network state $\bm\chi^1 = \bm\chi \triangleq (\bm\chi_n = (L_n, A_{n, (\mathrm{t})}, W_n): n \in \mathcal{N})$ can be expressed as in (\ref{expePayo}) on the bottom of Page \pageref{expePayo},
\begin{figure*}[!b]
\hrule
\begin{align}\label{expePayo}
  V_i\!\left(\bm\chi, \bm\pi\right) =
  \left(1 - \gamma\right) \cdot \textsf{E}_{\bm\pi}\!\!\left[\sum_{k = 1}^{\infty} (\gamma)^{k - 1} \cdot
  F_i\!\left(\bm\chi^k, \bm\varphi_i\!\left(\bm\pi_{(\mathrm{c})}\!\left(\bm\chi^k\right)\right), \bm\pi_{i, (\mathrm{t})}\!\left(\bm\chi_i^k\right),
  \bm\pi_{i, (\mathrm{p})}\!\left(\bm\chi_i^k\right)\right) | \bm\chi^1 = \bm\chi\right]
\end{align}
\end{figure*}
where $\gamma \in [0, 1)$ is a discount factor and $(\gamma)^{k-1}$ denotes the discount factor to the $(k-1)$-th power.
$V_i(\bm\chi, \bm\pi)$ is also termed as the state-value function of SP $i$ in a global network state $\bm\chi$ under a joint control policy $\bm\pi$.
The aim of each SP $i$ is to device a best-response control policy $\bm\pi_i^*$ that maximizes $V_i(\bm\chi, \bm\pi_i, \bm\pi_{-i})$ for any given initial network state $\bm\chi$, which can be formally formulated as
\begin{equation}\label{bestResp}
  \bm\pi_i^* = \underset{\bm\pi_i}{\arg\max}~ V_i(\bm\chi, \bm\pi_i, \bm\pi_{-i}), \forall \bm\chi \in \mathcal{X}^{|\mathcal{N}|}.
\end{equation}
A Nash equilibrium (NE) describes the rational behaviours of the SPs in a stochastic game.

\emph{Definition 1.}
In our formulated stochastic game, $\mathcal{SG}$, a NE is a tuple of control policies $\langle \bm\pi_i^*: i \in \mathcal{I}\rangle$, where each $\bm\pi_i^*$ of a SP $i$ is the best response to the other SPs' $\bm\pi_{-i}^*$.

For the $I$-player stochastic game $\mathcal{SG}$ with expected infinite-horizon discounted payoffs, there always exists a NE in stationary control policies \cite{Fink64}.
Define $\mathds{V}_i(\bm\chi) = V_i(\bm\chi, \bm\pi_i^*, \bm\pi_{-i}^*)$ as the optimal state-value function, $\forall i \in \mathcal{I}$ and $\forall \bm\chi \in \mathcal{X}^{|\mathcal{N}|}$.

\emph{Remark 1:}
From (\ref{expePayo}), we can easily observe that the expected long-term payoff of a SP $i \in \mathcal{I}$ depends on information of not only the global network state across time horizon but also the joint control policy $\bm\pi$.
In other words, the decision makings from all SPs are coupled in $\mathcal{SG}$.

\subsection{Best-Response Approach}

Suppose that in the $\mathcal{SG}$, the global network state information is known and all SPs play the NE control policies $\bm\pi^*$, the best-response of a SP $i \in \mathcal{I}$ under $\bm\chi \in \mathcal{X}^{|\mathcal{N}|}$ can then be obtained as (\ref{VFunc}) on the bottom of Page \pageref{VFunc},
\begin{figure*}[!b]
\hrule
\begin{align}\label{VFunc}
      \mathds{V}_i(\bm\chi) 
  & = \max\limits_{\bm\pi_i(\bm\chi)}\!\Bigg\{(1 - \gamma) \cdot 
            F_i\!\left(\bm\chi, \bm\varphi_i\!\left(\pi_{i, (\mathrm{c})}(\bm\chi), \bm\pi_{-i, (\mathrm{c})}^*(\bm\chi)\right),
            \bm\pi_{i, (\mathrm{t})}(\bm\chi_i), \bm\pi_{i, (\mathrm{p})}(\bm\chi_i)\right)                                                             \\
  & + \gamma \cdot \sum_{\bm\chi' \in \mathcal{X}^{|\mathcal{N}|}}
            \mathbb{P}\!\left(\bm\chi' | \bm\chi, \bm\varphi\!\left(\pi_{i, (\mathrm{c})}(\bm\chi), \bm\pi_{-i, (\mathrm{c})}^*(\bm\chi)\right),
            \left(\bm\pi_{i, (\mathrm{t})}(\bm\chi_i), \bm\pi_{-i, (\mathrm{t})}^*(\bm\chi_{-i})\right),
            \left(\bm\pi_{i, (\mathrm{p})}(\bm\chi_i), \bm\pi_{-i, (\mathrm{p})}^*(\bm\chi_{-i})\right)\right) \cdot \mathds{V}_i(\bm\chi') \Bigg\}     \nonumber
\end{align}
\end{figure*}
where $\bm\chi_i = (\bm\chi_n: n \in \mathcal{N}_i)$ and $\bm\chi' = (\bm\chi_n' = (L_n', A_{n, (\mathrm{t})}', W_n'): n \in \mathcal{N})$ is the next global network state.

\emph{Remark 2:}
It is a challenging task to find the NE for the $\mathcal{SG}$.
In order to operate the NE, all SPs have to know the global network dynamics, which is prohibited in our non-cooperative networking environment.

\section{Abstract Stochastic Game Reformulation and Deep Reinforcement Learning}
\label{probSolv}

In this section, we elaborate on how SPs play the cross-slice resource orchestration stochastic game with limited information.
We reformulate an abstract stochastic game with the conjectures of the interactions among the competing SPs.
By linearly decomposing the abstract state-value functions of a SP, we derive a DRL-based online learning scheme to approximate the optimal control policies.

\subsection{Stochastic Game Abstraction via Conjectures}
\label{SGA}

To capture the coupling of decision makings among the competing SPs, we abstract $\mathcal{SG}$ as $\mathcal{AG}$ \cite{Kroe16, Chen1801}.
In the abstract stochastic game $\mathcal{AG}$, a SP $i \in \mathcal{I}$ behaves based on its own local network dynamics and abstractions of states at other competing SPs.
Let $\mathcal{S}_i = \{1, \cdots, S_i\}$ be the abstraction of state space $\mathcal{X}_{-i}$, where $S_i \in \mathds{N}_+$.
The existing mechanisms for state abstraction are NP-complete \cite{Abel16} and require full network state information sharing among SPs.
On the other hand, we note that the behavioural couplings in $\mathcal{SG}$ exist in the channel auction \cite{Fu09} and the payments of SP $i$ depend on $\mathcal{X}_{-i}$.
We allow each SP $i$ in $\mathcal{AG}$ to construct $\mathcal{S}_i$ by classifying the value region $[0, \Gamma_i]$\footnote{From the analysis in previous sections, the payment by the SP $i$ to the SDN-orchestrator at a scheduling slot is of finite value.} of payments into $S_i$ intervals, namely, $[0, \Gamma_{i, 1}]$, $(\Gamma_{i, 1}, \Gamma_{i, 2}]$, $(\Gamma_{i, 2}, \Gamma_{i, 3}]$, $\ldots$, $(\Gamma_{i, S_i - 1}, \Gamma_{i, S_i}]$, where $\Gamma_{i, S_i} = \Gamma_i$ is the maximum payment value and we let $\Gamma_{i, 1} = 0$ for a special case in which SP $i$ wins the auction but pays nothing to the SDN-orchestrator\footnote{With the VCG mechanism, such a case exists when the number of channels in the network is sufficiently large \cite{Jia09}.}.
With this regard, a global network state $(\bm\chi_i, \bm\chi_{- i}) \in \mathcal{X}^{|\mathcal{N}|}$ is conjectured as $\tilde{\bm\chi}_i = (\bm\chi_i, s_i) \in \tilde{\mathcal{X}}_i$ if SP $i$ receives a payment $\tau_i$ in $(\Gamma_{i, s_i - 1}, \Gamma_{i, s_i}]$ from the channel auction in previous scheduling slot, where $\tilde{\mathcal{X}}_i = \mathcal{X}_i \times \mathcal{S}_i$ and $s_i \in \mathcal{S}_i$.
Hence $\mathcal{S}_i$ can be treated as an approximation of $\mathcal{X}_{-i}$ but with the size $S_i \ll |\mathcal{X}_{-i}|$.
To ease the analysis in the following, we mathematically represent the conjecture by a surjective mapping function $g_i: \mathcal{X}_{- i} \rightarrow \mathcal{S}_i$.

\emph{Remark 3:}
Classifying the payment values brings the immediate benefit of a much reduced abstract state space for a SP.
More importantly, the conjecture makes it possible for the prediction of expected future payment, which is needed when we specify an auction bid in the next Section \ref{VDeco}.

Let $\tilde{\bm\pi}_i = (\tilde{\pi}_{i, (\mathrm{c})}, \bm\pi_{i, (\mathrm{t})}, \bm\pi_{i, (\mathrm{p})})$ be the abstract control policy in the abstract stochastic game $\mathcal{AG}$ played by a SP $i \in \mathcal{I}$ over the abstract network state space $\tilde{\mathcal{X}}_i$, where $\tilde{\pi}_{i, (\mathrm{c})}$ is the abstract channel auction policy.
In the abstraction from stochastic game $\mathcal{SG}$ to $\mathcal{AG}$ for SP $i$, we have (\ref{payoGap}) on the bottom of Page \pageref{payoGap},
\begin{figure*}[!b]
\hrule
\begin{align}\label{payoGap}
  & \Gamma_{i, s_i} - \Gamma_{i, s_i - 1} \geq                                                                                              \nonumber\\
  & \max_{\{\bm\chi: g_i(\bm\chi_{- i}) = s_i\}}
    \left|
        F_i\!\left(\bm\chi, \bm\varphi_i\!\left(\bm\pi_{(\mathrm{c})}(\bm\chi)\right), \bm\pi_{i, (\mathrm{t})}(\bm\chi_i),
            \bm\pi_{i, (\mathrm{p})}(\bm\chi_i)\right) -
        \tilde{F}_i\!\left(\tilde{\bm\chi}_i, \bm\varphi_i\!\left(\tilde{\bm\pi}_{(\mathrm{c})}\!\left(\tilde{\bm\chi}\right)\right),
            \bm\pi_{i, (\mathrm{t})}(\bm\chi_i), \bm\pi_{i, (\mathrm{p})}(\bm\chi_i)\right)
    \right|
\end{align}
\end{figure*}
where $\tilde{F}_i(\tilde{\bm\chi}_i, \bm\varphi_i(\tilde{\bm\pi}_{(\mathrm{c})}(\tilde{\bm\chi})), \bm\pi_{i, (\mathrm{t})}(\bm\chi_i), \bm\pi_{i, (\mathrm{p})}(\bm\chi_i))$ is the payoff of SP $i$ in $\tilde{\bm\chi}_i \in \tilde{\mathcal{X}}_i$ under $\tilde{\bm\pi}_i$, $\tilde{\bm\chi} = (\tilde{\bm\chi}_i: i \in \mathcal{I})$, $\tilde{\bm\pi}_{(\mathrm{c})} = (\tilde{\pi}_{i, (\mathrm{c})}: i \in \mathcal{I})$, and $\bm\pi_{(\mathrm{c})}$ is the original joint channel auction policy in $\mathcal{SG}$.
Likewise, the abstract state-value function for SP $i$ under $\tilde{\bm\pi} = (\tilde{\bm\pi}_i: i \in \mathcal{I})$ can be defined as (\ref{abstVFunc}) on the bottom of Page \pageref{abstVFunc}, $\forall \tilde{\bm\chi}_i \in \tilde{\mathcal{X}}_i$,
\begin{figure*}[!b]
\hrule
\begin{align}\label{abstVFunc}
       \tilde{V}_i\!\left(\tilde{\bm\chi}_i, \tilde{\bm\pi}\right)
    = (1 - \gamma) \cdot \textsf{E}_{\tilde{\bm\pi}}\!\!\left[\sum_{k = 1}^{\infty} (\gamma)^{k - 1} \cdot
      \tilde{F}_i\!\left(\tilde{\bm\chi}_i^k, \bm\varphi_i\!\left(\tilde{\bm\pi}_{(\mathrm{c})}\!\left(\tilde{\bm\chi}^k\right)\right), \bm\pi_{i, (\mathrm{t})}\!\left(\bm\chi_i^k\right), \bm\pi_{i, (\mathrm{p})}\!\left(\bm\chi_i^k\right)\right) | \tilde{\bm\chi}_i^1 = \tilde{\bm\chi}_i\right]
\end{align}
\end{figure*}
where $\tilde{\bm\chi}^k = (\tilde{\bm\chi}_i^k = (\bm\chi_i^k, s_i^k): i \in \mathcal{I})$ with $s_i^k$ being the abstract state at slot $k$.
We will shortly see in Lemma 1 that the expected long-term payoff achieved by SP $i$ from the $\tilde{\bm\pi}_i$ in $\mathcal{AG}$ is not far from that from the original $\bm\pi_i$ in $\mathcal{SG}$.
Let $\Upsilon_i = \max_{s_i \in \mathcal{S}_i} (\Gamma_{i, s_i} - \Gamma_{i, s_i - 1})$.

\emph{Lemma 1:}
For an original control policy $\bm\pi$ and the corresponding abstract policy $\tilde{\bm\pi}$ in games $\mathcal{SG}$ and $\mathcal{AG}$, we have, $\forall i \in \mathcal{I}$, $|V_i(\bm\chi, \bm\pi) - \tilde{V}_i(\tilde{\bm\chi}_i, \tilde{\bm\pi})| \leq \Upsilon_i$, $\forall \bm\chi \in \mathcal{X}^{|\mathcal{N}|}$, where $\tilde{\bm\chi}_i = (\bm\chi_i, s_i)$ with $s_i = g_i(\bm\chi_{-i})$.

\emph{Proof:}
The proof proceeds similar to \cite{Chen1801, Abel16}.
\hfill $\Box$

Instead of playing the original joint control policy $\bm\pi^*$ in the stochastic game $\mathcal{SG}$, Theorem 1 shows that the NE joint abstract control policy given by $\tilde{\bm\pi}^* = (\tilde{\bm\pi}_i^*: i \in \mathcal{I})$ in the abstract stochastic game $\mathcal{AG}$ leads to a bounded regret, where $\tilde{\bm\pi}_i^* = (\tilde{\pi}_{i, (\mathrm{c})}^*, \bm\pi_{i, (\mathrm{t})}^*, \bm\pi_{i, (\mathrm{p})}^*)$ denotes the best-response abstract control policy of SP $i \in \mathcal{I}$.

\emph{Theorem 1.}
For a SP $i \in \mathcal{I}$, let $\bm\pi_i$ be the original control policy corresponding to an abstract control policy $\tilde{\bm\pi}_i$.
The original joint control policy $\bm\pi^*$ corresponding to a joint abstract control policy $\tilde{\bm\pi}^*$ satisfies $V_i(\bm\chi, (\bm\pi_i, \bm\pi_{-i}^*)) \leq \mathds{V}_i(\bm\chi) + 2 \cdot \Upsilon_i$, $\forall \bm\chi \in \mathcal{X}^{|\mathcal{N}|}$, where $(\bm\pi_i, \bm\pi_{-i}^*)$ is the joint control policy that results from SP $i$ unilaterally deviating from $\bm\pi_i^*$ to $\bm\pi_i$ in the original stochastic game $\mathcal{SG}$.

\emph{Proof:}
The proof uses a contradiction.
Assume for a SP $i \in \mathcal{I}$, there exists $\bm\chi \in \mathcal{X}^{|\mathcal{N}|}$ such that $V_i(\bm\chi, (\bm\pi_i, \bm\pi_{- i}^*)) > \mathds{V}_i(\bm\chi) + 2 \cdot \Upsilon_i$, where $\bm\pi_i$ is the original control policy corresponding to a non-best-response abstract control policy $\tilde{\bm\pi}_i$.
Using the result from Lemma 1, we arrive at
\begin{align}
        \tilde{V}_i\!\left(\tilde{\bm\chi}_i, \left(\tilde{\bm\pi}_i, \tilde{\bm\pi}_{- i}^*\right)\right)
 & \geq V_i\!\left(\bm\chi, \left(\bm\pi_i, \bm\pi_{- i}^*\right)\right) - \Upsilon_i                                                               \nonumber\\
 & >    \left(\mathds{V}_i\!\left(\bm\chi\right) + 2 \cdot \Upsilon_i\right) - \Upsilon_i                                                           \nonumber\\
 & \geq \left(\left(\tilde{V}_i\!\left(\tilde{\bm\chi}_i, \tilde{\bm\pi}^*\right) - \Upsilon_i\right) + 2 \cdot \Upsilon_i\right) - \Upsilon_i      \nonumber\\
 & =    \tilde{V}_i\!\left(\tilde{\bm\chi}_i, \tilde{\bm\pi}^*\right),
\end{align}
which is contradicted the definition of a NE in the abstract stochastic game $\mathcal{AG}$.
This concludes the proof. \hfill $\Box$

Hereinafter, we switch our focus from the stochastic game $\mathcal{SG}$ to the abstract stochastic game $\mathcal{AG}$.
Suppose all SPs play the NE joint abstract control policy $\tilde{\bm\pi}^*$ in the abstract stochastic game $\mathcal{AG}$.
Denote $\tilde{\mathds{V}}_i(\tilde{\bm\chi}_i) = \tilde{V}_i(\tilde{\bm\chi}_i, \tilde{\bm\pi}^*)$, $\forall \tilde{\bm\chi}_i \in \tilde{\mathcal{X}}_i$ and $\forall i \in \mathcal{I}$.
The best-response abstract control policy of a SP $i$ can be computed as in (\ref{ASVF}) on the bottom of Page \pageref{ASVF},
\begin{figure*}[!b]
\hrule
\begin{align}\label{ASVF}
     \tilde{\mathds{V}}_i\!\left(\tilde{\bm\chi}_i\right)
 & = \max_{\tilde{\bm\pi}_i(\tilde{\bm\chi}_i)}\! \Bigg\{(1 - \gamma) \cdot
     \tilde{F}_i\!\left(\tilde{\bm\chi}_i, \bm\varphi_i\!\left(\tilde{\pi}_{i, (\mathrm{c})}\!\left(\tilde{\bm\chi}_i\right),
     \tilde{\bm\pi}_{-i, (\mathrm{c})}^*(\tilde{\bm\chi}_{-i})\right),
     \bm\pi_{i, (\mathrm{t})}(\bm\chi_i), \bm\pi_{i, (\mathrm{p})}(\bm\chi_i)\right)                                                                \nonumber\\
 & + \gamma \cdot \sum_{\tilde{\bm\chi}_i' \in \tilde{\mathcal{X}}_i} \mathbb{P}\!\left(\tilde{\bm\chi}_i' | \tilde{\bm\chi}_i,
     \bm\varphi_i\!\left(\tilde{\pi}_{i, (\mathrm{c})}\!\left(\tilde{\bm\chi}_i\right),
     \tilde{\bm\pi}_{-i, (\mathrm{c})}^*(\tilde{\bm\chi}_{-i})\right), \bm\pi_{i, (\mathrm{t})}(\bm\chi_i), \bm\pi_{i, (\mathrm{p})}(\bm\chi_i)\right) \cdot
     \tilde{\mathds{V}}_i\!\left(\tilde{\bm\chi}_i'\right)\!\Bigg\}
\end{align}
\end{figure*}
$\forall \tilde{\bm\chi}_i \in \tilde{\mathcal{X}}_i$, which is based on only the local information.

\emph{Remark 4:}
There remain two challenges involved in solving (\ref{ASVF}) for each SP $i \in \mathcal{I}$: 1) a priori knowledge of the abstract network state transition probability, which incorporates the statistics of MU mobilities, the computation task and packet arrivals and the conjectures of other competing SPs' local network information (i.e., the statistics of $\mathcal{S}_i$), is not feasible; and 2) given a specific classification of the payment values, the size of the decision making space $\{\tilde{\bm\pi}_i(\tilde{\bm\chi}_i): \tilde{\bm\chi}_i \in \tilde{\mathcal{X}}_i\}$ grows exponentially as $|\mathcal{N}_i|$ increases.

\subsection{Decomposition of Abstract State-Value Function}
\label{VDeco}

Observing that: 1) the channel auction decision as well as the computation offloading and packet scheduling decisions are made in sequence and are independent across a SP and its subscribed MUs; and 2) the per-slot instantaneous payoff function (\ref{payOff}) of a SP is of an additive nature, we are hence motivated to decompose the per-SP MDP described by (\ref{ASVF}) into $|\mathcal{N}_i| + 1$ independent single-agent MDPs.
More specifically, for a SP $i \in \mathcal{I}$, the abstract state-value function $\tilde{\mathds{V}}_i(\tilde{\bm\chi}_i)$, $\forall \tilde{\bm\chi}_i \in \tilde{\mathcal{X}}_i$, can be calculated as

\begin{equation}\label{VFactDeco}
   \tilde{\mathds{V}}_i\!\left(\tilde{\bm\chi}_i\right)
 = \sum\limits_{n \in \mathcal{N}_i} \alpha_n \cdot \mathds{U}_n(\bm\chi_n) - \mathds{U}_i(s_i),
\end{equation}
where the per-MU expected long-term utility $\mathds{U}_n(\bm\chi_n)$ and the expected long-term payment $\mathds{U}_i(s_i)$ of SP $i$ satisfy, respectively,
\begin{align}\label{MTBell}
 &   \mathds{U}_n(\bm\chi_n) = \\
 & \max_{R_{n, (\mathrm{t})}, R_{n, (\mathrm{p})}} \!\!\Bigg\{\!(1 - \gamma) \!\cdot\!
     U_n\!\left(\!\bm\chi_n, \varphi_n\!\!\left(\tilde{\bm\pi}_{(\mathrm{c})}^*(\tilde{\bm\chi})\right)\!,
     R_{n, (\mathrm{t})}, R_{n, (\mathrm{p})}\right)\! +                                                                                    \nonumber\\
 & \gamma \!\cdot\! \sum_{\bm\chi_n' \in \mathcal{X}}
     \mathbb{P}\!\left(\!\bm\chi_n' | \bm\chi_n, \varphi_n\!\left(\tilde{\bm\pi}_{(\mathrm{c})}^*\!\left(\tilde{\bm\chi}\right)\right)\!,
     R_{n, (\mathrm{t})}, R_{n, (\mathrm{p})}\right) \!\cdot\! \mathds{U}_n\!\left(\bm\chi_n'\right)\!\Bigg\},\nonumber
\end{align}
and
\begin{align}\label{WSPBell}
     \mathds{U}_i(s_i)
 & = (1 - \gamma) \cdot \tau_i                                                                                                        \nonumber\\
 & + \gamma \cdot \sum_{s_i' \in \mathcal{S}_i} \mathbb{P}\!\left(s_i' | s_i,
     \phi_i\!\left(\tilde{\bm\pi}_{(\mathrm{c})}^*\!\left(\tilde{\bm\chi}\right)\right)\right) \cdot \mathds{U}_i\!\left(s_i'\right),
\end{align}
with $\tilde{\bm\pi}_{(\mathrm{c})}^*(\tilde{\bm\chi}) = (\tilde{\pi}_{i, (\mathrm{c})}^*(\tilde{\bm\chi}_i): i \in \mathcal{I})$, while $R_{n, (\mathrm{t})}$ and $R_{n, (\mathrm{p})}$ being the computation offloading and packet scheduling decisions under a current local network state $\bm\chi_n$ of MU $n \in \mathcal{N}_i$.
It is worth to note that the winner determination and the payment calculation from the VCG auction at the SDN-orchestrator deduce the derivation of (\ref{WSPBell}).

\emph{Remark 5:}
We highlight below two key advantages of the linear decomposition approach in (\ref{VFactDeco}).
\begin{enumerate}
  \item Simplified decision makings: The linear decomposition motivates a SP $i \in \mathcal{I}$ to let the MUs locally make the computation offloading and packet scheduling decisions, which reduces the action space of size $(\mathcal{A} \times \mathcal{W})^{|\mathcal{N}_i|}$ at SP $i$ to $|\mathcal{N}_i|$ local spaces of size $\mathcal{A} \times \mathcal{W}$ at the MUs.
  \item Near optimality: The linear decomposition approach, which can be viewed as a special case of the feature-based decomposition method \cite{Tsit96}, provides an accuracy guarantee of the approximation of the abstract state-value function \cite{Rich98}.
\end{enumerate}

With the decomposition of the abstract state-value function as in (\ref{VFactDeco}), we can now specify the number of requested channels by a SP $i \in \mathcal{I}$ in the coverage of a BS $b \in \mathcal{B}$ as
\begin{align}\label{numChan}
  C_{b, i} = \sum\limits_{\left\{n \in \mathcal{N}_i: L_n \in \mathcal{L}_b\right\}} z_n,
\end{align}
and the true value of obtaining $\mathbf{C}_i = (C_{b, i}: b \in \mathcal{B})$ across the service region as
\begin{align}\label{valu}
      \nu_i
  & = \frac{1}{1 - \gamma} \cdot \sum\limits_{n \in \mathcal{N}_i} \alpha_n \cdot \mathds{U}_n(\bm\chi_n)                   \nonumber\\
  & - \frac{\gamma}{1 - \gamma} \cdot \sum_{s_i' \in \mathcal{S}_i}
      \mathbb{P}\!\left(s_i' | s_i, \mathds{1}_{\left\{\sum_{b \in \mathcal{B}} C_{b, i} > 0\right\}}\right) \cdot
      \mathds{U}_i\!\left(s_i'\right),
\end{align}
which together constitute the optimal bid $\tilde{\pi}_{i, (\mathrm{c})}^*(\tilde{\bm\chi}_i) = \bm\beta_i \triangleq (\nu_i, \mathbf{C}_i)$ of SP $i$ under a current abstract network state $\tilde{\bm\chi}_i \in \tilde{\mathcal{X}}_i$, where for a MU $n \in \mathcal{N}_i$, $z_n$ given by
\begin{align}\label{whetChan}
 & z_n =                                                                                                                                    \\
 & \underset{z \in \{0, 1\}}{\arg\max} \Bigg\{(1 - \gamma) \cdot
     U_n\!\left(\bm\chi_n, z, \pi_{n, (\mathrm{t})}^*(\bm\chi_n), \pi_{n, (\mathrm{p})}^*(\bm\chi_n)\right) +                               \nonumber\\
 & \gamma \cdot \sum_{\bm\chi_n' \in \mathcal{X}}
     \mathbb{P}\!\left(\!\bm\chi_n' | \bm\chi_n, z, \pi_{n, (\mathrm{t})}^*(\bm\chi_n), \pi_{n, (\mathrm{p})}^*(\bm\chi_n)\right) \cdot
     \mathds{U}_n(\bm\chi_n')\Bigg\},                                                                                                       \nonumber
\end{align}
indicates the preference of obtaining one channel, and $\mathds{1}_{\{\Xi\}}$ is an indicator function that equals $1$ if the condition $\Xi$ is satisfied and $0$ otherwise.
We can easily find that the calculation of the optimal bid $\bm\beta_i$ at SP $i$ needs the private information of $(s_i, \mathbb{P}(s' | s, \iota - 1))$ and $(\mathds{U}_n(\bm\chi_n), z_n, L_n)$ from each subscribed MU $n \in \mathcal{N}_i$, where $s' \in \mathcal{S}_i$ and $\iota \in \{1, 2\}$.

\subsection{Learning Optimal Abstract Control Policy}
\label{loacp}

In the calculation of true value as in (\ref{valu}) for a SP $i \in \mathcal{I}$ at the beginning of each scheduling slot $k$, the abstract network state transition probability $\mathbb{P}(s' | s, \iota - 1)$, which is necessary for the prediction of the value of expected future payments, is unknown.
We propose that SP $i$ maintains over the scheduling slots a three-dimensional table $\mathbf{Y}_i^k$ of size $S_i \cdot S_i \cdot 2$.
Each entry $y_{s, s', \iota}^k$ in table $\mathbf{Y}_i^k$ represents the number of transitions from $s_i^{k - 1} = s$ to $s_i^k = s'$ when $\phi_i^{k - 1} = \iota - 1$ up to scheduling slot $k$.
$\mathbf{Y}_i^k$ is updated using the channel auction outcomes from the SDN-orchestrator.
Then, the abstract network state transition probability at scheduling slot $k$ can be estimated to be\footnote{To ensure that division by zero is not possible, each entry in a table $\mathbf{Y}_i^1$, $\forall i \in \mathcal{I}$, needs to be initialized, for example, to 1 as in numerical experiments.}
\begin{align}
    \mathbb{P}\!\left(s_i^k = s' | s_i^{k - 1} = s, \phi_i^{k - 1} = \iota - 1\right) =
    \frac{y_{s, s', \iota}^k}{\sum\limits_{s'' \in \mathcal{S}_i} y_{s'', s', \iota}^k}.
\end{align}
Applying the union bound and the weak law of large numbers \cite{Loev77}, (\ref{probConv}) (which is shown on the bottom of Page \pageref{probConv}) establishes
\begin{figure*}[!b]
\hrule
\begin{align}\label{probConv}
   \lim_{k \rightarrow\infty} \mathbb{P}\!\left(\left|
   \mathbb{P}\!\left(s_i^{k + 1} = s' | s_i^k = s, \phi_i^k = \iota - 1\right)
 - \mathbb{P}\!\left(s_i^k = s' | s_i^{k - 1} = s, \phi_i^{k - 1} = \iota - 1\right)\right| > \omega \right) = 0
\end{align}
\end{figure*}
for an arbitrarily small constant $\omega \in \mathds{R}_+$, $\forall s$, $s' \in \mathcal{S}_i$ and $\forall \iota \in \{1, 2\}$.
The state-value function $\mathds{U}_i(s_i)$, $\forall s_i \in \mathcal{S}_i$, is learned according to (\ref{paymUpda}) (which is shown on the bottom of Page \pageref{paymUpda})
\begin{figure*}[!b]
\hrule
\begin{align}\label{paymUpda}
   \mathds{U}_i^{k + 1}(s_i) =
   \left\{\!\!
   \begin{array}{l@{~}l}
     \left(1 - \zeta^k\right) \cdot \mathds{U}_i^k(s_i) + \zeta^k \cdot \left((1 - \gamma) \cdot \tau_i^k +
            \gamma \cdot \displaystyle\sum_{s_i^{k + 1} \in \mathcal{S}_i} \mathbb{P}\!\left(s_i^{k + 1} | s_i, \phi_i^k\right) \cdot
            \mathds{U}_i^k\!\left(s_i^{k + 1}\right)\right),                                                    & \mbox{if } s_i = s_i^k   \\
     \mathds{U}_i^k(s_i),                                                                                       & \mbox{otherwise}
   \end{array}
   \right.
\end{align}
\end{figure*}
based on $\phi_i^k$ and $\tau_i^k$ from the channel auction, where $\zeta^k \in [0, 1)$ is the learning rate.
The convergence of (\ref{paymUpda}) is guaranteed by $\sum_{k = 1}^\infty \zeta^k = \infty$ and $\sum_{k = 1}^\infty (\zeta^k)^2 < \infty$ \cite{Rich98}.

Given that all SPs deploy the best-response channel auction policies, the well-known value iteration \cite{Rich98} can be used by the MUs to find the optimal per-MU state-value functions (\ref{MTBell}).
However, this method requires full knowledge of the local network state transition probabilities, which is challenging without a priori statistical information of MU mobility, computation task arrivals and packet arrivals.

\subsubsection{Conventional $Q$-learning}
One attractiveness of the $Q$-learning is that it assumes no a priori knowledge of the local network state transition statistics.
Combining (\ref{MTBell}) and (\ref{whetChan}), we define for each MU $n \in \mathcal{N}$ the optimal state action-value function $Q_n: \mathcal{X} \times \{0, 1\} \times \mathcal{A} \times \mathcal{W} \rightarrow \mathds{R}$,
\begin{align}\label{QFunc}
 &   Q_n\!\left(\bm\chi_n, \varphi_n, R_{n, (\mathrm{t})}, R_{n, (\mathrm{p})}\right)                                                                   \nonumber\\
 & = (1 - \gamma) \cdot U_n\!\left(\bm\chi_n, \varphi_n, R_{n, (\mathrm{t})}, R_{n, (\mathrm{p})}\right)                                                \nonumber\\
 & + \gamma \cdot \sum_{\bm\chi_n' \in \mathcal{X}}
     \mathbb{P}\!\left(\bm\chi_n' | \bm\chi_n, \varphi_n, R_{n, (\mathrm{t})}, R_{n, (\mathrm{p})}\right) \cdot \mathds{U}_n\!\left(\bm\chi_n'\right),
\end{align}
where an action $(\varphi_n, R_{n, (\mathrm{t})}, R_{n, (\mathrm{p})})$ under a current local network state $\bm\chi_n$ consists of the channel allocation, computation offloading and packet scheduling decisions.
The optimal state-value function $\mathds{U}_n(\bm\chi_n)$ can be hence derived from
\begin{align}\label{stat_acti_q2}
  \mathds{U}_n(\bm\chi_n) =
  \max_{\varphi_n, R_{n, (\mathrm{t})}, R_{n, (\mathrm{p})}} Q_n\!\left(\bm\chi_n, \varphi_n, R_{n, (\mathrm{t})}, R_{n, (\mathrm{p})}\right).
\end{align}
By substituting (\ref{stat_acti_q2}) into (\ref{QFunc}), we get (\ref{MTBellQFunc}) on the bottom of Page \pageref{MTBellQFunc},
\begin{figure*}[!b]
\hrule
\begin{align}\label{MTBellQFunc}
     Q_n\!\left(\bm\chi_n, \varphi_n, R_{n, (\mathrm{t})}, R_{n, (\mathrm{p})}\right)
 & = (1 - \gamma) \cdot U_n\!\left(\bm\chi_n, \varphi_n, R_{n, (\mathrm{t})}, R_{n, (\mathrm{p})}\right)          \nonumber\\
 & + \gamma \cdot \sum_{\bm\chi_n' \in \mathcal{X}}
     \mathbb{P}\!\left(\bm\chi_n' | \bm\chi_n, \varphi_n, R_{n, (\mathrm{t})}, R_{n, (\mathrm{p})}\right) \cdot
     \max_{\varphi_n', R_{n, (\mathrm{t})}', R_{n, (\mathrm{p})}'}
     Q_n\!\left(\bm\chi_n', \varphi_n', R_{n, (\mathrm{t})}', R_{n, (\mathrm{p})}'\right)
\end{align}
\end{figure*}
where $(\varphi_n', R_{n, (\mathrm{t})}',$ $R_{n, (\mathrm{p})}')$ is an action under $\bm\chi_n'$.
Using $Q$-learning, the MU finds $Q_n(\bm\chi_n, \varphi_n, R_{n, (\mathrm{t})}, R_{n, (\mathrm{p})})$ iteratively using observations of the local network state $\bm\chi_n = \bm\chi_n^k$ at a current scheduling slot $k$, the action $(\varphi_n, R_{n, (\mathrm{t})}, R_{n, (\mathrm{p})}) = (\varphi_n^k, R_{n, (\mathrm{t})}^k, R_{n, (\mathrm{p})}^k)$, the achieved utility $U_n(\bm\chi_n, \varphi_n, R_{n, (\mathrm{t})}, R_{n, (\mathrm{p})})$ and the resulting local network state $\bm\chi_n' = \bm\chi_n^{k + 1}$ at the next slot $k + 1$.
The learning rule is given in (\ref{QLearRule}) on the bottom of Page \pageref{QLearRule},
\begin{figure*}[!b]
\hrule
\begin{align}\label{QLearRule}
 &   Q_n^{k + 1}\!\left(\bm\chi_n, \varphi_n, R_{n, (\mathrm{t})}, R_{n, (\mathrm{p})}\right)
   = Q_n^k\!\left(\bm\chi_n, \varphi_n, R_{n, (\mathrm{t})}, R_{n, (\mathrm{p})}\right) +                                   \\
 &   \zeta^k \cdot \left((1 - \gamma) U_n\!\left(\bm\chi_n, \varphi_n, R_{n, (\mathrm{t})}, R_{n, (\mathrm{p})}\right)
   + \gamma \cdot \max_{\varphi_n', R_{n, (\mathrm{t})}', R_{n, (\mathrm{p})}'}
     Q_n^k\!\left(\bm\chi_n', \varphi_n', R_{n, (\mathrm{t})}', R_{n, (\mathrm{p})}'\right)
   - Q_n^k\!\left(\bm\chi_n, \varphi_n, R_{n, (\mathrm{t})}, R_{n, (\mathrm{p})}\right)\right)                              \nonumber
\end{align}
\end{figure*}
which converges to the optimal control policy if: a) the local network state transition probability is stationary; and b) all state-action pairs are visited infinitely often \cite{Watk12}.
Condition b) can be satisfied when the probability of choosing any action in any local network state is non-zero (i.e., \emph{exploration}).
Meanwhile, in order to behave well, a MU has to exploit the most recently learned $Q$-function (i.e., \emph{exploitation}).
A classical way to balance \emph{exploration} and \emph{exploitation} is the $\epsilon$-greedy strategy \cite{Rich98}.

\textit{Remark 6:}
The tabular nature in representing $Q$-function values makes the conventional $Q$-learning not readily applicable to high-dimensional scenarios with huge state space, where the learning process can be extremely slow.
In our system, the sizes of local network state space $\mathcal{X}$ and action space $\{0, 1\} \times$ $\mathcal{A} \times \mathcal{W}$ are calculated as $|\mathcal{L}| \cdot (1 + A_{(\mathrm{t})}^{(\max)}) \cdot (1 + W^{(\max)})$ and $2 \cdot (1 + A_{(\mathrm{t})}^{(\max)}) \cdot (1 + W^{(\max)})$, respectively.
Consider a service region of $1.6 \cdot 10^3$ locations (as the network simulated in \cite{Chen15} and the numerical experiments in Section \ref{simu}), $A_{(\mathrm{t})}^{(\max)} = 5$ and $W^{(\max)} = 10$, the MU has to update totally $1.39392 \cdot 10^7$ $Q$-function values, which is impossible for the conventional $Q$-learning process to converge within a limited number of scheduling slots.

\subsubsection{Deep Reinforcement Learning}
\begin{figure*}[t]
  \centering
  \includegraphics[width=36.9pc]{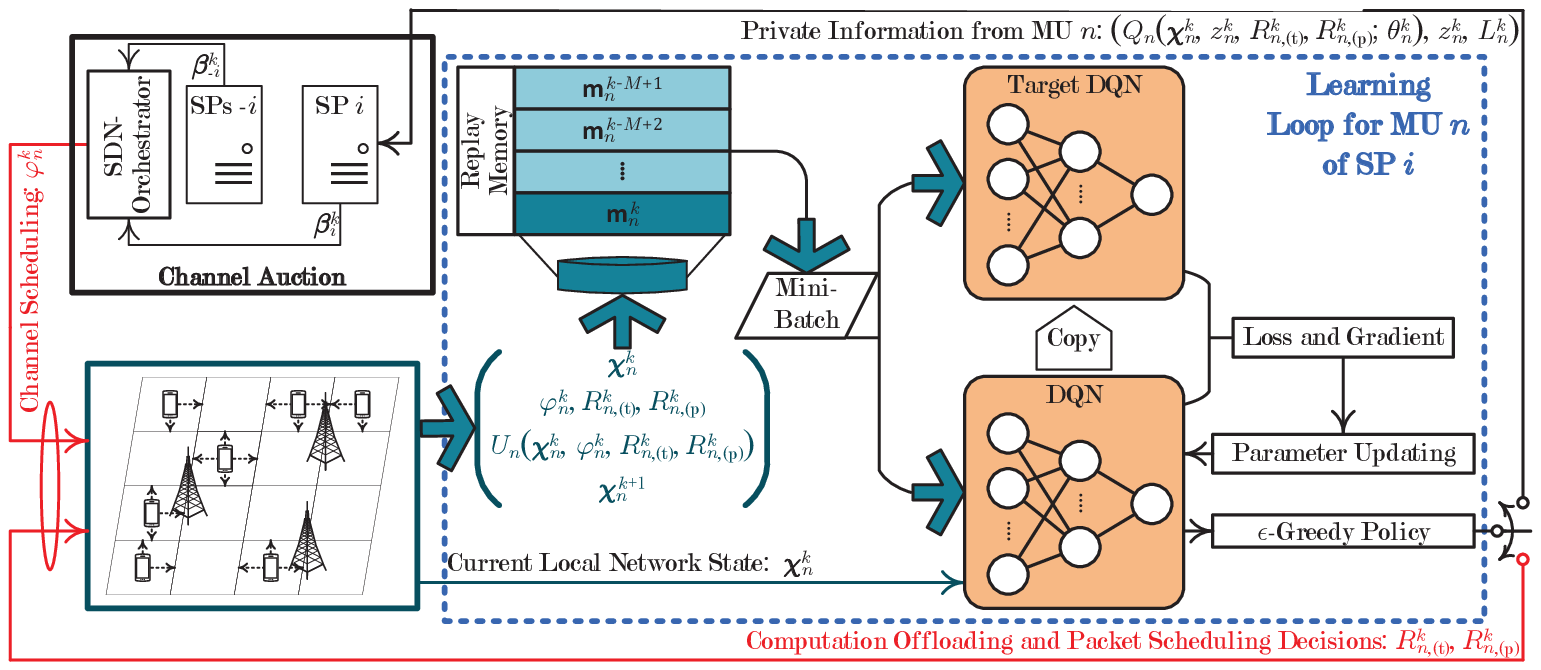}
  \caption{Application of a double deep $Q$-network (DQN) to approximate the optimal state action-value $Q$-function of a mobile user (MU) $n \in \mathcal{N}_i$ subscribed to service provider (SP) $i \in \mathcal{I}$.}
  \label{deepLear}
\end{figure*}

The advances in neural networks \cite{Appl17} and the success of a deep neural network in modelling an optimal state-action $Q$-function \cite{Mnih15} inspire us to resort to a double deep $Q$-network (DQN) to address the massive local network state space $\mathcal{X}$ at each MU $n \in \mathcal{N}_i$ of a SP $i \in \mathcal{I}$ in our considered system \cite{Hass16}.
That is, the $Q$-function in (\ref{MTBellQFunc}) can be approximated by $Q_n(\bm\chi_n, \varphi_n, R_{n, (\mathrm{t})}, R_{n, (\mathrm{p})}) \approx Q_n(\bm\chi_n, \varphi_n, R_{n, (\mathrm{t})}, R_{n, (\mathrm{p})}; \bm\theta_n)$, where $\bm\theta_n$ denotes a vector of parameters associated with the DQN of MU $n$.
The implementation of such a DRL algorithm for finding the approximated $Q$-function of MU $n$ is illustrated in Fig. \ref{deepLear}.

More specifically, each MU $n \in \mathcal{N}_i$ of a SP $i \in \mathcal{I}$ is equipped with a replay memory of a finite size $M$ to store the experience $\mathbf{m}_n^k$ given by
\begin{align}\label{exper}
 & \mathbf{m}_n^k =\\
 & \left(\!\bm\chi_n^k, \left(\!\varphi_n^k, R_{n, (\mathrm{t})}^k, R_{n, (\mathrm{p})}^k\!\right)\!,
        U_n\!\!\left(\!\bm\chi_n^k, \varphi_n^k, R_{n, (\mathrm{t})}^k, R_{n, (\mathrm{p})}^k\!\right)\!, \bm\chi_n^{k + 1}\right)\!, \nonumber
\end{align}
which is happened at the transition between two consecutive scheduling slots $k$ and $k + 1$ during the process of DRL.
The memory of experiences can be encapsulated as $\mathcal{M}_n^k = \{\mathbf{m}_n^{k - M + 1}, \cdots, \mathbf{m}_n^k\}$.
Each MU $n$ maintains a DQN as well as a target DQN, namely, $Q_n(\bm\chi_n, \varphi_n, R_{n, (\mathrm{t})}, R_{n, (\mathrm{p})}; \bm\theta_n^k)$ and $Q_n(\bm\chi_n, \varphi_n, R_{n, (\mathrm{t})}, R_{n, (\mathrm{p})}; \bm\theta_{n, -}^k)$, with $\bm\theta_n^k$ and $\bm\theta_{n, -}^k$ being the associated parameters at a current scheduling slot $k$ and a certain previous scheduling slot before slot $k$, respectively.
According to the experience replay technique \cite{Lin92}, at each scheduling slot $k$, MU $n$ randomly samples a mini-batch $\mathcal{O}_n^k \subseteq \mathcal{M}_n^k$ of size $O < M$ from the replay memory $\mathcal{M}_n^k$ to train the DQN.
The training objective is to update the parameters $\bm\theta_n^k$ in the direction of minimizing the loss function given by (\ref{lossFunc}) on the bottom of Page \pageref{lossFunc},
\begin{figure*}[!b]
\hrule
\begin{align}\label{lossFunc}
 &   \textsf{LOSS}_n\!\left(\bm\theta_n^k\right)
   = \textsf{E}_{\left(\bm\chi_n, (\varphi_n, R_{n, (\mathrm{t})}, R_{n, (\mathrm{p})}),
     U_n(\bm\chi_n, \varphi_n, R_{n, (\mathrm{t})}, R_{n, (\mathrm{p})}), \bm\chi_n'\right) \in \mathcal{O}_n^k}\Bigg[\Bigg(
     (1 - \gamma) \cdot U_n(\bm\chi_n, \varphi_n, R_{n, (\mathrm{t})}, R_{n, (\mathrm{p})})~+                                       \nonumber\\
 &   \gamma \cdot Q_n\!\left(\bm\chi_n', \underset{\varphi_n', R_{n, (\mathrm{t})}',
     R_{n, (\mathrm{p})}'}{\arg\max} Q_n\!\left(\bm\chi_n', \varphi_n', R_{n, (\mathrm{t})}',
     R_{n, (\mathrm{p})}'; \bm\theta_n^k\right); \bm\theta_{n, -}^k\right) -
     Q_n\!\left(\bm\chi_n, \varphi_n, R_{n, (\mathrm{t})}, R_{n, (\mathrm{p})}; \bm\theta_n^k\right)\Bigg)^2\Bigg]
\end{align}
\end{figure*}
which is a mean-squared measure of the Bellman equation error at a scheduling slot $k$.
By differentiating $\textsf{LOSS}_n(\bm\theta_n^k)$ with respect to $\bm\theta_n^k$, we obtain the gradient as in (\ref{grad}) on the bottom of Page \pageref{grad}.
\begin{figure*}[!b]
\hrule
\begin{align}\label{grad}
 &   \nabla_{\bm\theta_n^k} \textsf{LOSS}_n\!\left(\bm\theta_n^k\right) =
     \textsf{E}_{\left(\bm\chi_n, (\varphi_n, R_{n, (\mathrm{t})}, R_{n, (\mathrm{p})}),
     U_n(\bm\chi_n, \varphi_n, R_{n, (\mathrm{t})}, R_{n, (\mathrm{p})}), \bm\chi_n'\right) \in \mathcal{O}_n^k}\Bigg[
     \Bigg((1 - \gamma) \cdot U_n(\bm\chi_n, \varphi_n, R_{n, (\mathrm{t})}, R_{n, (\mathrm{p})})~+                                         \nonumber\\
 &   \gamma \cdot Q_n\!\left(\bm\chi_n', \underset{\varphi_n', R_{n, (\mathrm{t})}', R_{n, (\mathrm{p})}'}{\arg\max}~
     Q_n\!\left(\bm\chi_n', \varphi_n', R_{n, (\mathrm{t})}', R_{n, (\mathrm{p})}'; \bm\theta_n^k\right); \bm\theta_{n, -}^k\right) -
     Q_n\!\left(\bm\chi_n, \varphi_n, R_{n, (\mathrm{t})}, R_{n, (\mathrm{p})}; \bm\theta_n^k\right)\Bigg) \cdot                            \nonumber\\
 & \nabla_{\bm\theta_n^k} Q_n\!\left(\bm\chi_n, \varphi_n, R_{n, (\mathrm{t})}, R_{n, (\mathrm{p})}; \bm\theta_n^k\right)\Bigg]
\end{align}
\end{figure*}
Algorithm \ref{algo} details the online training procedure of MU $n$.

\begin{algorithm}[t]
    \caption{Online DRL for Approximating Optimal State Action-Value $Q$-Functions of a MU $n \in \mathcal{N}_i$ of a SP $i \in \mathcal{I}$}
    \label{algo}
    \begin{algorithmic}[1]
        \STATE \textbf{initialize} the replay memory $\mathcal{M}_n^k$ of size $M \in \mathds{N}_+$, the mini-batch $\mathcal{O}_n^k$ of size $O < M$, a DQN and a target DQN with two sets $\bm\theta_n^k$ and $\bm\theta_{n, -}^k$ of parameters, and the local network state $\bm\chi_n^k$, for $k = 1$.

        \REPEAT
            \STATE At the beginning of scheduling slot $k$, the MU observes the packet arrivals $A_{n, (\mathrm{p})}^k$, takes $\bm\chi_n^k$ as an input to the DQN with parameters $\bm\theta_n^k$, and then selects a random action $(z_n^k, R_{n, (\mathrm{t})}^k, R_{n, (\mathrm{p})}^k)$ with probability $\epsilon$ or with probability $1 - \epsilon$, an action $(z_n^k, R_{n, (\mathrm{t})}^k, R_{n, (\mathrm{p})}^k)$ that is with maximum value $Q_n(\bm\chi_n^k, z_n^k, R_{n, (\mathrm{t})}^k, R_{n, (\mathrm{p})}^k; \bm\theta_n^k)$.

            \STATE MU $n$ sends $[Q_n(\bm\chi_n^k, z_n^k, R_{n, (\mathrm{t})}^k, R_{n, (\mathrm{p})}^k; \bm\theta_n^k), z_n^k, L_n^k)$ to the subscribing SP $i$. SP $i$ submits its bidding vector $\bm\beta_i = (\nu_i, \mathbf{C}_i)$ to the SDN-orchestrator, where $\nu_i$ is given by (\ref{valu}) and $\mathbf{C}_i = (C_{b, i}: b \in \mathcal{B})$ with each $C_{b, i}$ given by (\ref{numChan}).

            \STATE With the bids from all SPs, the SDN-orchestrator determines the auction winners $\bm\phi^k$ and channel allocation $\bm\rho_i^k = (\bm\rho_n^k: n \in \mathcal{N}_i)$ according to (\ref{chanSche}), and calculates the payments $\tau_i^k$ according to (\ref{paymCalc}) for SP $i$.

            \STATE With the channel allocation $\bm\rho_n^k$, winner determination $\phi_i^k$ and payment $\tau_i^k$, SP $i$ updates $\mathbf{Y}_i^k$ and $\mathds{U}_i^{k + 1}(s_i^k)$ according to (\ref{paymUpda}), and MU $n$ makes computation offloading $\varphi_n^k R_{n, (\mathrm{t})}^k$ and packet scheduling $\varphi_n^k R_{n, (\mathrm{p})}^k$.

            \STATE MU $n$ achieves utility $U_n(\bm\chi_n^k, \varphi_n^k, R_{n, (\mathrm{t})}^k, R_{n, (\mathrm{p})}^k)$ and observes $\bm\chi_n^{k + 1}$ at the next slot $k + 1$.

            \STATE MU $n$ updates the $\mathcal{M}_n^k$ with $\mathbf{m}_n^k = (\bm\chi_n^k, (\varphi_n^k, R_{n, (\mathrm{t})}^k,$ $R_{n, (\mathrm{p})}^k), U_n(\bm\chi_n^k, \varphi_n^k, R_{n, (\mathrm{t})}^k, R_{n, (\mathrm{p})}^k), \bm\chi_n^{k + 1})$.

            \STATE With a randomly sampled $\mathcal{O}_n^k$ from $\mathcal{M}_n^k$, MU $n$ updates the DQN parameters $\bm\theta_n^k$ with the gradient in (\ref{grad}).

            \STATE MU $n$ regularly reset the target DQN parameters with $\bm\theta_{n, -}^{k + 1} = \bm\theta_n^k$, and otherwise $\bm\theta_{n, -}^{k + 1} = \bm\theta_{n, -}^k$.

            \STATE The scheduling slot index is updated by $k \leftarrow k + 1$.
        \UNTIL{A predefined stopping condition is satisfied.}
    \end{algorithmic}
\end{algorithm}

\section{Numerical Experiments}
\label{simu}

In order to quantify the performance gain from the proposed DRL-based online learning scheme for multi-tenant cross-slice resource orchestration in a software-defined RAN, numerical experiments based on TensorFlow \cite{Abad16} are conducted.

\subsection{Parameter Settings}

\begin{figure}[t]
    \centering
    \includegraphics[width=19.6pc]{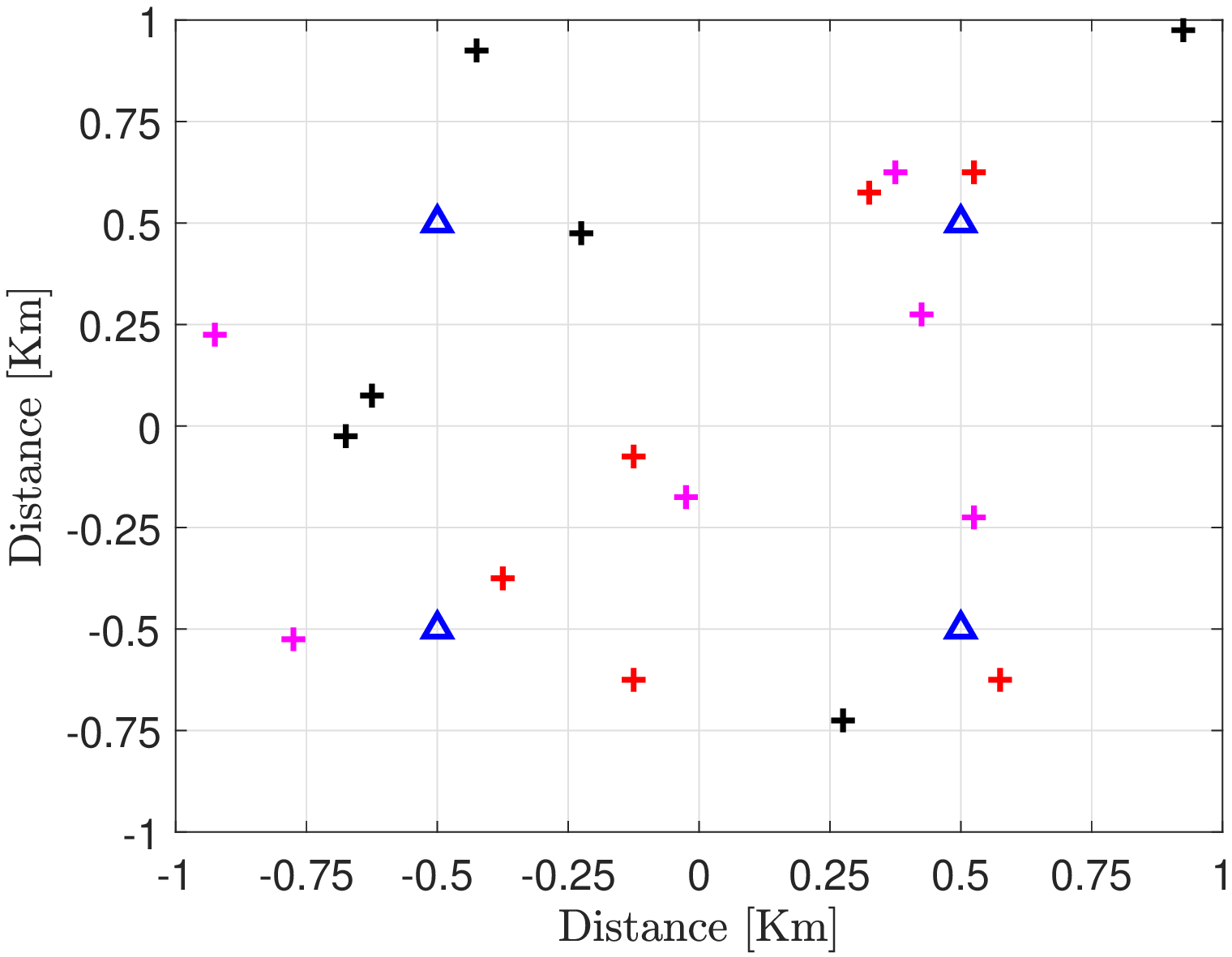}
    \caption{A snapshot of the RAN: BSs, MUs of SPs 1, 2 and 3 are shown in blue triangles, red, black and magenta pluses, respectively.}
    \label{netLayout}
\end{figure}

For experimental purpose, we build up a physical RAN, which is composed of $4$ BSs in a $2\times2$ Km$^2$ square area.
Fig. \ref{netLayout} shows the layout of the RAN.
The BSs are placed at equal distance apart.
The entire service region is divided into $1600$ locations with each representing a small area of 50$\times$50 m$^2$.
In other words, each BS covers $400$ locations\footnote{The numerical experiments can be readily extended to other network layouts, such as \cite{Siom12}.}.
The average channel gain experienced by a MU $n \in \mathcal{N}$ in the coverage of a BS $b \in \mathcal{B}$ at location $L_n^k \in \mathcal{L}_b$ during a scheduling slot $k$ is given as \cite{Mao16}
\begin{align}
    h(L_n^k) = H_0 \cdot \left(\frac{\xi_0}{\xi_{b, n}^k}\right)^4,
\end{align}
where $H_0 = -40$ dB is the path-loss constant, $\xi_0 = 2$ m is the reference distance and $\xi_{b, n}^k$ is the physical distance between MU $n$ and BS $b$.
The state transition probability matrices for the Markov processes of mobilities and computation task arrivals of all MUs are independently and randomly generated.
The packet arrivals are assumed to follow a Poisson arrival process with average rate $\lambda$ (in packets/slot).
$U_n^{(1)}(\cdot)$, $U_n^{(2)}(\cdot)$, $U_n^{(3)}(\cdot)$ and $U_n^{(4)}(\cdot)$ in (\ref{util}) are chosen to be
\begin{align}
    U_n^{(1)}\!\left(W_n^{k + 1}\right)             & = \exp\!\left\{- W_n^{k + 1}\right\},                         \\
    U_n^{(2)}\!\left(D_n^k\right)                   & = \exp\!\left\{- D_n^k\right\},                               \\
    U_n^{(3)}\!\left(P_{n, (\mathrm{CPU})}^k\right) & = \exp\!\left\{- P_{n, (\mathrm{CPU})}^k\right\},             \\
    U_n^{(4)}\!\left(P_{n, (\mathrm{tr})}^k\right)  & = \exp\!\left\{- P_{n, (\mathrm{tr})}^k\right\}.
\end{align}
For a MU, we design a DQN with $2$ hidden layers with each consisting of $16$ neurons\footnote{The tradeoff between the time spent during the training process and the performance improvement with a deeper and/or wider neural network is still an open problem \cite{CCLP18, Chen1802}.}.
Tanh is selected as the activation function \cite{Jarr09} and Adam as the optimizer \cite{King15}.
With the consideration of limited memory capacity of mobile devices, we set the replay size as $M = 5000$.
Other parameter values used in the experiments are listed in Table \ref{tabl2}.

\begin{table}[t]
  \caption{Parameter values in experiments.}\label{tabl2}
        \begin{center}
        \begin{tabular}{c|c}
              \hline
              Parameter                                                     & Value                                \\\hline
              Set of SPs                      $\mathcal{I}$                 & $\{1, 2, 3\}$                        \\\hline
              Set of BSs                      $\mathcal{B}$                 & $\{1, 2, 3, 4\}$                     \\\hline
              Number of MUs                   $|\mathcal{N}_i|$             & $6$, $\forall i \in\mathcal{I}$      \\\hline
              Channel bandwidth               $\eta$                        & $500$ KHz                            \\\hline
              Noise power spectral density    $\sigma^2$                    & $-174$ dBm/Hz                        \\\hline
              Scheduling slot duration        $\delta$                      & $10^{-2}$ second                     \\\hline
              Discount factor                 $\gamma$                      & $0.9$                                \\\hline
              Utility price                   $\alpha_n$                    & $1$, $\forall n\in\mathcal{N}$       \\\hline
              Packet size                     $\mu_{(\mathrm{p})}$          & $3000$ bits                          \\\hline
              Maximum transmit power          $\Omega^{(\max)}$                  & $3$ Watts                            \\\hline
              Weight of energy consumption    $\ell_n$                      & $3$, $\forall n\in\mathcal{N}$       \\\hline
              Maximum queue length            $W^{(\max)}$                  & $10$ packets                         \\\hline
              Maximum task arrivals           $A_{(\mathrm{t})}^{(\max)}$   & $5$ tasks                            \\\hline
              Input data size                 $\mu_{(\mathrm{t})}$          & $5000$ bits                          \\\hline
              CPU cycles per bit              $\vartheta$                   & $737.5$                              \\\hline
              CPU-cycle frequency             $\varrho$                     & $2$ GHz                              \\\hline
              Effective switched capacitance  $\varsigma$                   & $2.5 \cdot 10^{-28}$                 \\\hline
              Exploration probability         $\epsilon$                    & $0.001$                              \\
              \hline
        \end{tabular}
    \end{center}
\end{table}

For performance comparisons, three baseline schemes are simulated, namely,
\begin{enumerate}
  \item Channel-aware control policy (Baseline 1) -- At the beginning of each slot, the need of getting one channel at a MU is evaluated by the average channel gain;
  \item Queue-aware control policy (Baseline 2) -- Each MU calculates the preference between having one channel or not using a predefined threshold of the data queue length;
  \item Random control policy (Baseline 3) -- This policy randomly generates the value of obtaining one channel for each MU at each scheduling slot.
\end{enumerate}
During the implementation of the above three baselines, after the centralized channel allocation at the SDN-orchestrator, each MU proceeds to select a random number of computation tasks for offloading and decides a maximum feasible number of packets for transmission \cite{Fu09}.

\subsection{Experiment Results}

\subsubsection{Experiment 1 -- Convergence performance}

\begin{figure}[t]
    \centering
    \includegraphics[width=19.6pc]{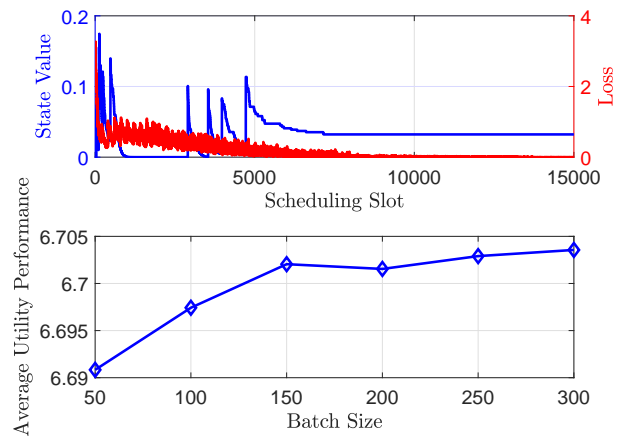}
    \caption{Illustration for convergence speed of the proposed online resource orchestration scheme based on DRL (upper) and average utility performance per MU across the learning procedure versus batch sizes (lower).}
    \label{conver}
\end{figure}

Our goal in this experiment is to validate if the considered system remains stable by implementing our proposed online learning scheme for multi-tenant cross-slice resource orchestration.
We fix the average packet arrival rate and the number of channels to be $\lambda = 6$ packets per slot and $J = 9$, respectively.
The batch size is set as $O = 200$.
In all experiments in this work, we use $S_i = 36$, $\forall i \in \{1, 2, 3\}$.
Without loss of the generality, we plot the variations in $\mathds{U}_1(2)$ of SP $1$ and the loss function $\textsf{LOSS}_1(\bm\theta_1^k)$ of a MU $1 \in \mathcal{N}_1 = \{1, 2, \cdots, 6\}$ versus the scheduling slots in the upper subplot in Fig. \ref{conver}, which validates the convergence behaviour of our scheme.
The learning scheme converges within $10^4$ scheduling slots.
In the lower subplot in Fig. \ref{conver}, we plot the average utility performance per MU with different batch size choices.
The average performance per MU across the learning procedure has been commonly selected as the metric, for example, in works \cite{Fu13, Chen1801}.
It is obvious from (\ref{grad}) that for a MU, a larger batch size results in a more stable gradient estimate, i.e., a smaller variance, hence a better average utility performance across the learning procedure.
Given the replay memory capacity, the utility performance improvement, however, saturates, when the batch size exceeds $150$.
Hence we continue to use $O = 200$ in Experiments 2 and 3 to strike a balance between performance improvement and computation overhead.

\subsubsection{Experiment 2 -- Performance under various $\lambda$}

\begin{figure}[t]
    \centering
    \includegraphics[width=19.6pc]{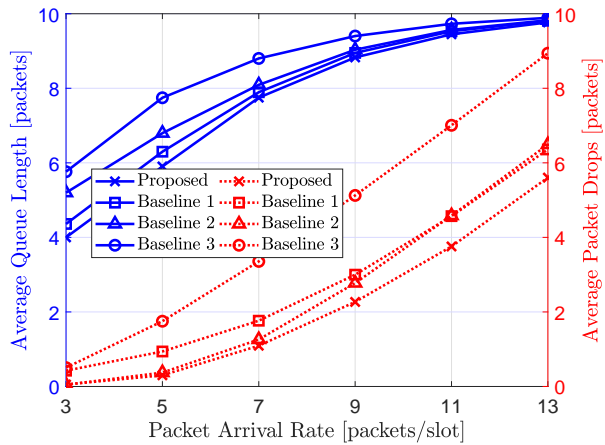}
    \caption{Average queue length and packet drops per MU across the learning procedure versus average packet arrival rates.}
    \label{simu01_01}
\end{figure}

\begin{figure}[t]
    \centering
    \includegraphics[width=19.6pc]{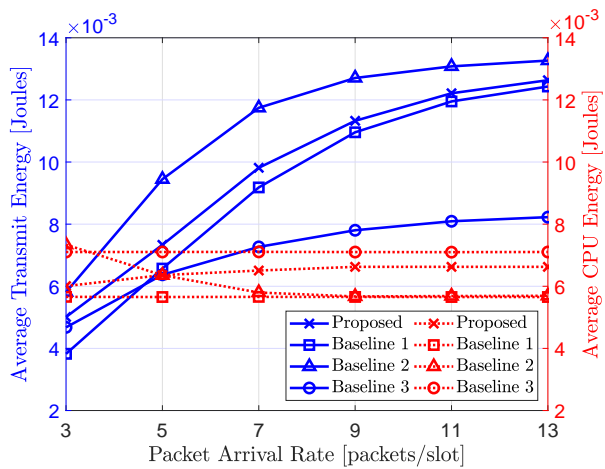}
    \caption{Average CPU energy and transmit energy consumptions per MU across the learning procedure versus average packet arrival rates.}
    \label{simu01_02}
\end{figure}

\begin{figure}[t]
    \centering
    \includegraphics[width=19.6pc]{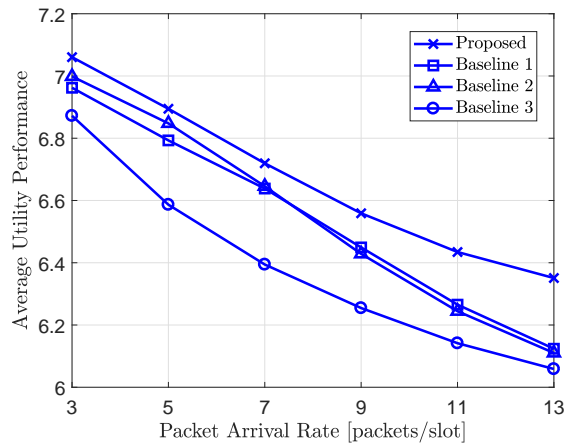}
    \caption{Average utility performance per MU across the learning procedure versus average packet arrival rates.}
    \label{simu01_03}
\end{figure}

This experiment primarily aims to demonstrate the average performance per scheduling slot in terms of the average queue length, the average packet drops, the average CPU energy consumption, the average transmit energy consumption and the average utility under different packet arrival rates.
We assume in the system $J = 11$ channels that can be used across the MUs to access the computation and communication slices.
The results are exhibited in Figs. \ref{simu01_01}, \ref{simu01_02} and \ref{simu01_03}.
Fig. \ref{simu01_01} illustrates the average queue length and the average packet drops per MU.
Fig. \ref{simu01_02} illustrates the average CPU energy consumption and the average transmit energy consumption per MU.
Fig. \ref{simu01_03} illustrates the average utility per MU.

Each plot compares the performance of our proposed scheme with the three baseline multi-tenant cross-slice resource orchestration schemes.
From Fig. \ref{simu01_03}, it can be observed that the proposed scheme achieves a significant gain in average utility per MU.
Similar observations can be made from the curves in Fig. \ref{simu01_01}, which shows that the minimum queue length and packet drops can be realized from the proposed scheme.
As the packet arrival rate increases, each MU consumes more transmit energy for the delivery of incoming data packets.
However, when implementing Baselines 1 and 3, the average CPU energy consumption per MU keeps constant due to the fact that the opportunities of winning the channel auction do not change.
On the other hand, the average CPU energy consumption per MU from Baseline 2 decreases since a larger queue length indicates a bigger chance of getting one channel and hence a higher probability of offloading the computation task.
In contrast to Baseline 2, the proposed scheme transmits more data packets to avoid packet drops by leaving more computation tasks processed at the mobile devices, leading to increased average CPU energy consumption.

\subsubsection{Experiment 3 -- Performance with different $J$}

\begin{figure}[t]
    \centering
    \includegraphics[width=19.6pc]{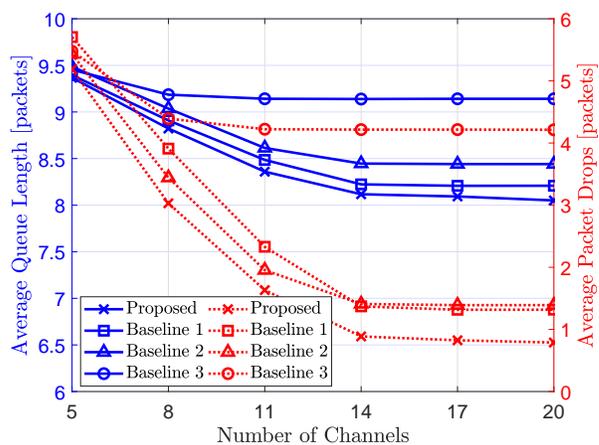}
    \caption{Average queue length and packet drops per MU across the learning procedure versus numbers of channels.}
    \label{simu02_01}
\end{figure}

\begin{figure}[t]
    \centering
    \includegraphics[width=19.6pc]{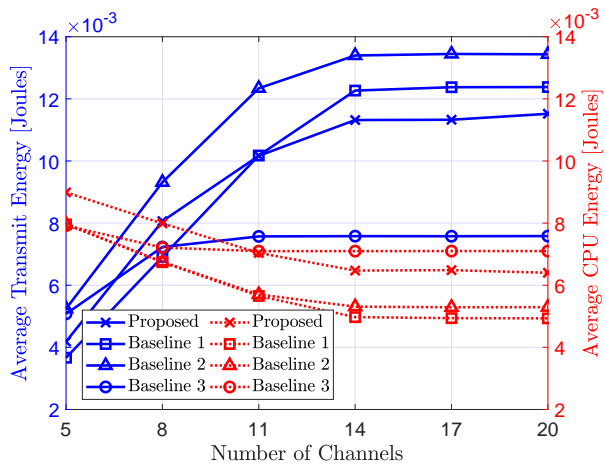}
    \caption{Average CPU energy and transmit energy consumptions per MU across the learning procedure versus numbers of channels.}
    \label{simu02_02}
\end{figure}

\begin{figure}[t]
    \centering
    \includegraphics[width=19.6pc]{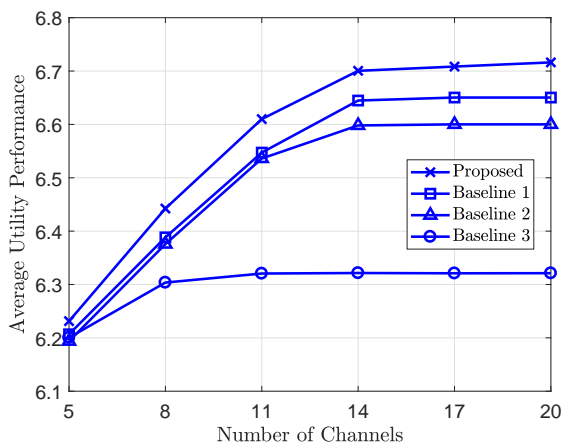}
    \caption{Average utility performance per MU across the learning procedure versus numbers of channels.}
    \label{simu02_03}
\end{figure}

In the last experiment, we simulate the average resource orchestration performance per scheduling slot achieved from the proposed scheme and other three baselines versus the numbers of channels.
The packet arrival rate in this experiment is selected as $\lambda = 8$.
The average queue length, average packet drops, average CPU energy consumption, average transmit energy consumption and average utility per MU across the entire learning period are depicted in Figs. \ref{simu02_01}, \ref{simu02_02} and \ref{simu02_03}.
It can be easily observed from Figs. \ref{simu02_01} and \ref{simu02_03} that as the number of available channels increases, the average queue length and the average packet drops decrease, while the average utility per MU improves.
As the number of channels that can be allocated to the MUs increases, it becomes more likely for a MU to obtain one channel.
Therefore, the MU is able to offload more computation tasks and transmit more data packets, while the average CPU energy consumption decreases and the average transmit energy consumption increases, as shown in Fig. \ref{simu02_02}.
From both Experiments 2 and 3, the proposed online learning scheme outperforms the three baselines.

\section{Conclusions}
\label{conc}

We study in this paper the multi-tenant cross-slice resource orchestration in a system with RAN-only slicing.
Over the scheduling slots, the competing SPs bid to orchestrate the limited channel access opportunities over their MUs with MEC and traditional mobile service requests to the computation and communication slices.
The SDN-orchestrator regulates the channel auction through a VCG pricing mechanism at the beginning of each slot.
We formulate the non-cooperative problem as a stochastic game, in which each SP aims to maximize its own expected long-term payoff.
However, the channel auction, computation offloading and packet scheduling decisions of a SP require complete information of the network dynamics as well as the control policies of other SPs.
To solve the problem, we approximate the interactions among the SPs by an abstract stochastic game.
In the abstract stochastic game, a SP is thus able to behave independently with the conjectures of other SPs' behaviours.
We observe that the channel auction decision and the computation offloading and packet scheduling decisions are sequentially made.
This motivates us to linearly decompose the per-SP MDP, which greatly simplifies the decision making process of a SP.
Furthermore, an online scheme based on DRL is proposed in order to find the optimal abstract control policies.
Numerical experiments showcase that significant performance gains can be achieved from our scheme, compared with the other baselines.

\section*{Acknowledgements}

The work carried out in this paper was supported by the Academy of Finland under Grants 319759, 319758 and 289611, the National Key R\&D Program of China under Grant 2017YFB1301003, the National Natural Science Foundation of China under Grants 61701439 and 61731002, the Zhejiang Key Research and Development Plan under Grant 2019C01002, the JSPS KAKENHI under Grant 18KK0279, and the Telecommunications Advanced Foundation.
The authors would like to sincerely thank the anonymous reviewers for their valuable comments which led to a significant improvement of this paper.

%
\begin{IEEEbiography}[{\includegraphics[width=1in, height=1.25in, clip]{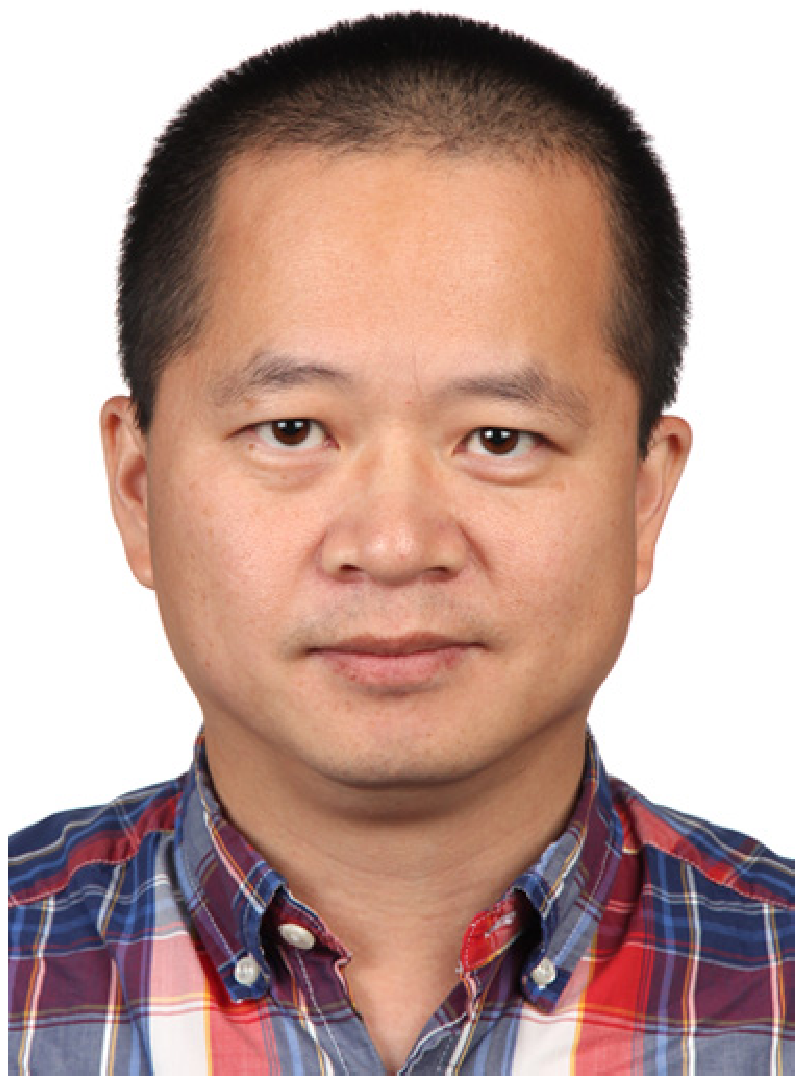}}]{Xianfu Chen}
received his Ph.D. degree with honors in Signal and Information Processing, from the Department of Information Science and Electronic Engineering (ISEE) at Zhejiang University, Hangzhou, China, in March 2012.
Since April 2012, Dr. Chen has been with the VTT Technical Research Centre of Finland, Oulu, Finland, where he is currently a Senior Scientist.
His research interests cover various aspects of wireless communications and networking, with emphasis on human-level and artificial intelligence for resource awareness in next-generation communication networks.
Dr. Chen is serving and served as a Track Co-Chair and a TPC member for a number of IEEE ComSoc flagship conferences.
He is a Vice Chair of IEEE Special Interest Group on Big Data with Computational Intelligence, the members of which come from over 15 countries worldwide.
He is an IEEE member.
\end{IEEEbiography}

\begin{IEEEbiography}[{\includegraphics[width=1in, height=1.25in, clip]{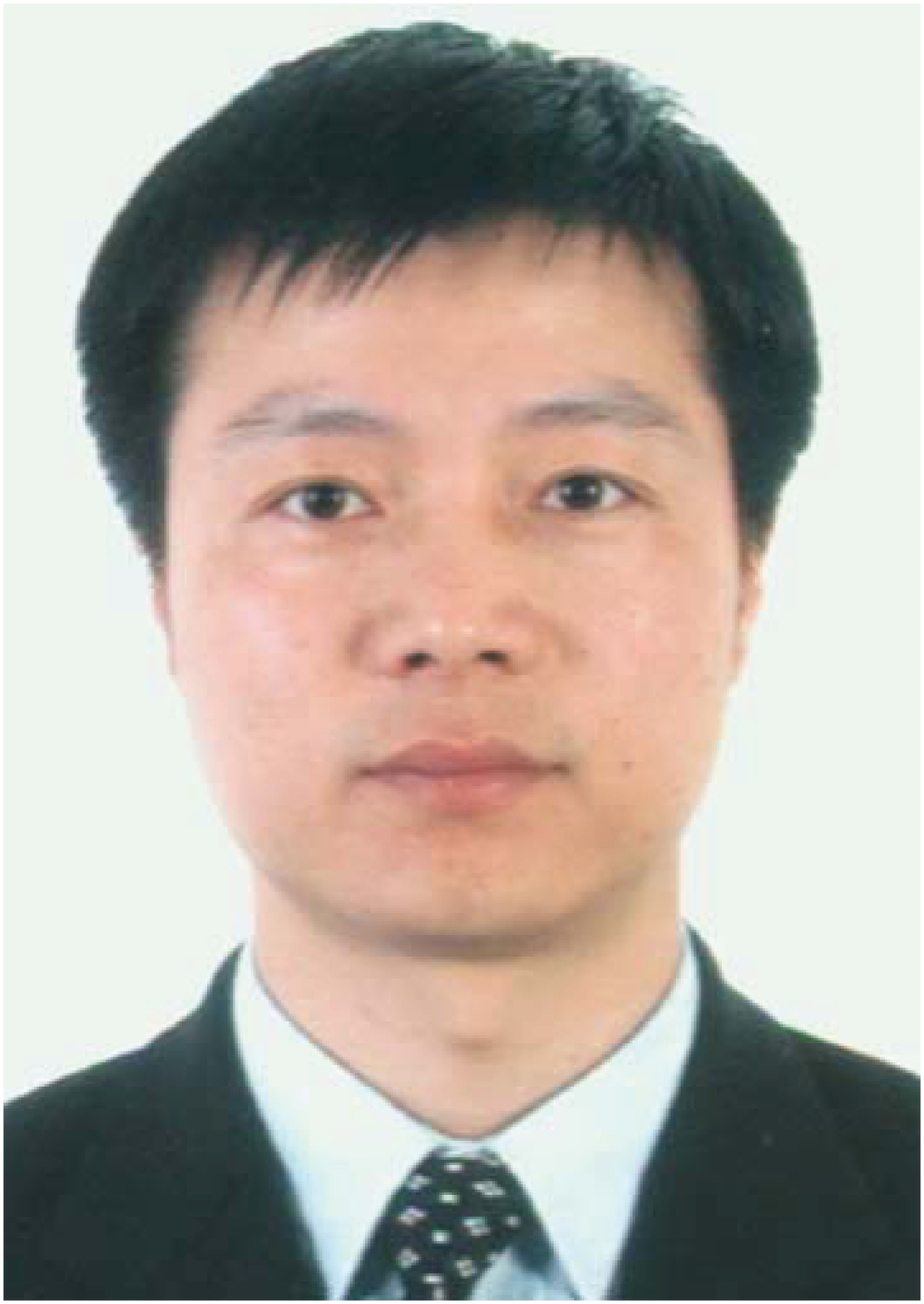}}]{Zhifeng Zhao}
received the Ph.D. degree in Communication and Information System from the PLA University of Science and Technology, Nanjing, China, in 2002.
From 2002 to 2004, he was a Postdoctoral Researcher with Zhejiang University, China, where his works were focused on multimedia next generation networks and soft-switch technology for energy efficiency.
From 2005 to 2006, he was a Senior Researcher with the PLA University of Science and Technology, where he performed research and development on advanced energy-efficient wireless router, ad hoc network simulator, and cognitive mesh networking test bed.
He is currently the Director of the Research Development Department at Zhejiang Lab, and is also an Associate Professor with the College of Information Science and Electronic Engineering at Zhejiang University, Hangzhou, China.
His research interests include cognitive radio, wireless multi-hop networks (ad hoc, mesh and wireless sensor networks), wireless multimedia networks, and green communications.
Dr. Zhao was the Symposium Co-Chair of the ChinaCom 2009 and 2010, and the TPC Co-Chair of the IEEE ISCIT 2010.
He is an IEEE member.
\end{IEEEbiography}

\begin{IEEEbiography}[{\includegraphics[width=1in, height=1.25in, clip, trim=10 0 10 5]{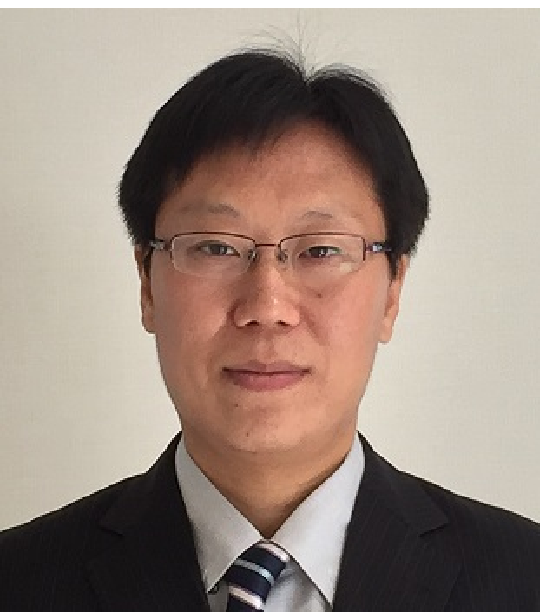}}]{Celimuge Wu}
received his ME degree from the Beijing Institute of Technology, China, in 2006, and his Ph.D. degree from The University of Electro-Communications, Japan, in 2010.
Dr. Wu is currently an associate professor with the Graduate School of Informatics and Engineering, The University of Electro-Communications.
His current research interests include vehicular networks, sensor networks, intelligent transport systems, IoT, and edge computing.
Dr. Wu is an associate editor of IEEE Access, IEICE Transactions on Communications, International Journal of Distributed Sensor Networks, and MDPI Sensors.
He is an IEEE senior member.
\end{IEEEbiography}

\begin{IEEEbiography}[{\includegraphics[width=1in, height=1.25in, clip, trim=60 0 60 0]{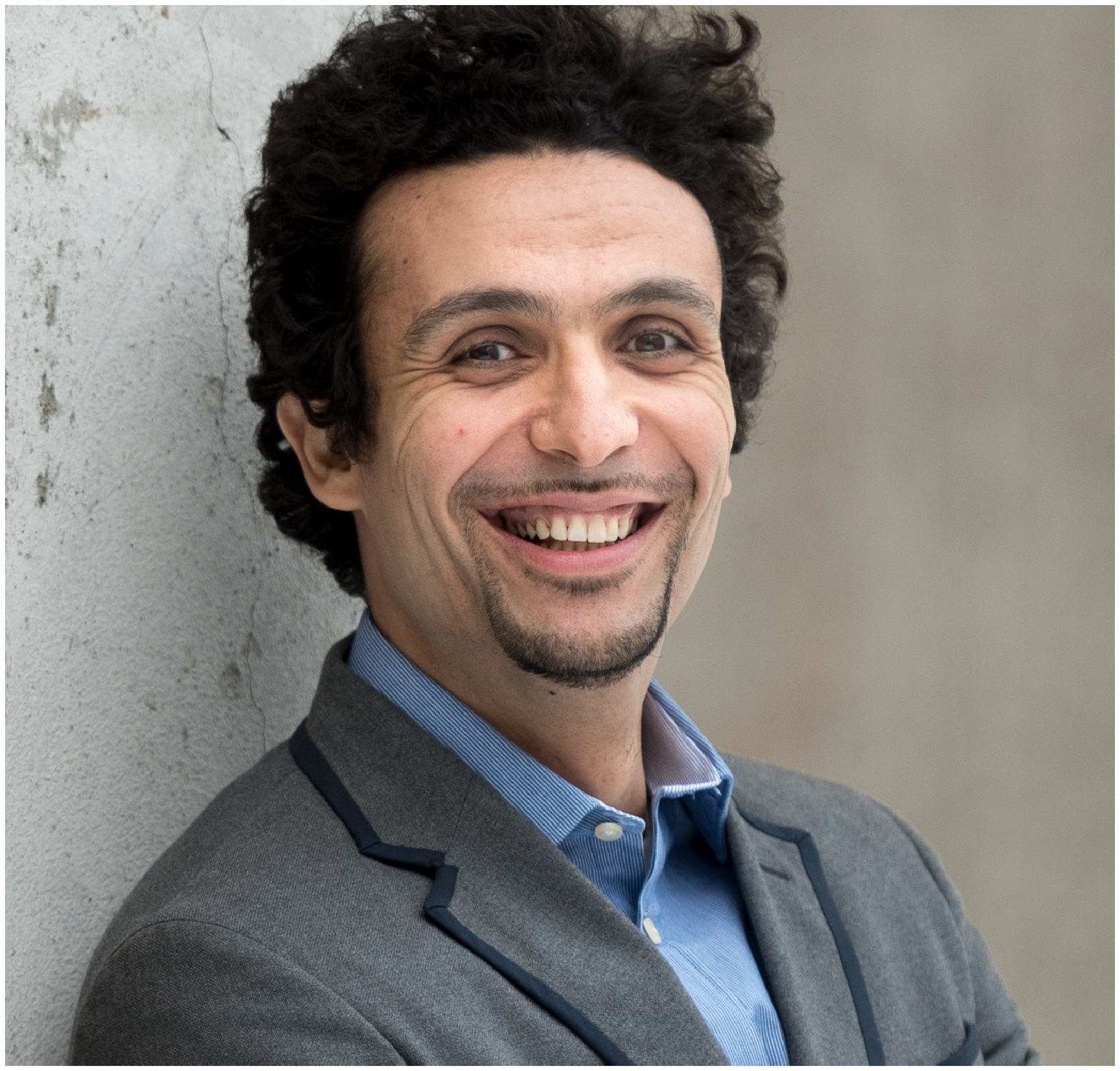}}]{Mehdi Bennis}
is an Associate Professor at the Centre for Wireless Communications, University of Oulu, Finland and an Academy of Finland Research Fellow.
His main research interests are in radio resource management, heterogeneous networks, game theory and machine learning in 5G networks and beyond.
He has co-authored one book and published more than 200 research papers in international conferences, journals and book chapters.
He has been the recipient of several awards, including the 2015 Fred W. Ellersick Prize from the IEEE Communications Society, the 2016 Best Tutorial Prize from the IEEE Communications Society, the 2017 EURASIP Best paper Award for the Journal of Wireless Communications and Networks, the all-University of Oulu award for research and the 2019 IEEE ComSoc Radio Communications Committee Early Achievement Award.
Dr. Bennis is an editor of IEEE Transactions on Communications.
He is an IEEE senior member.
\end{IEEEbiography}

\begin{IEEEbiography}[{\includegraphics[width=1in, height=1.25in, clip, trim=60 0 60 0]{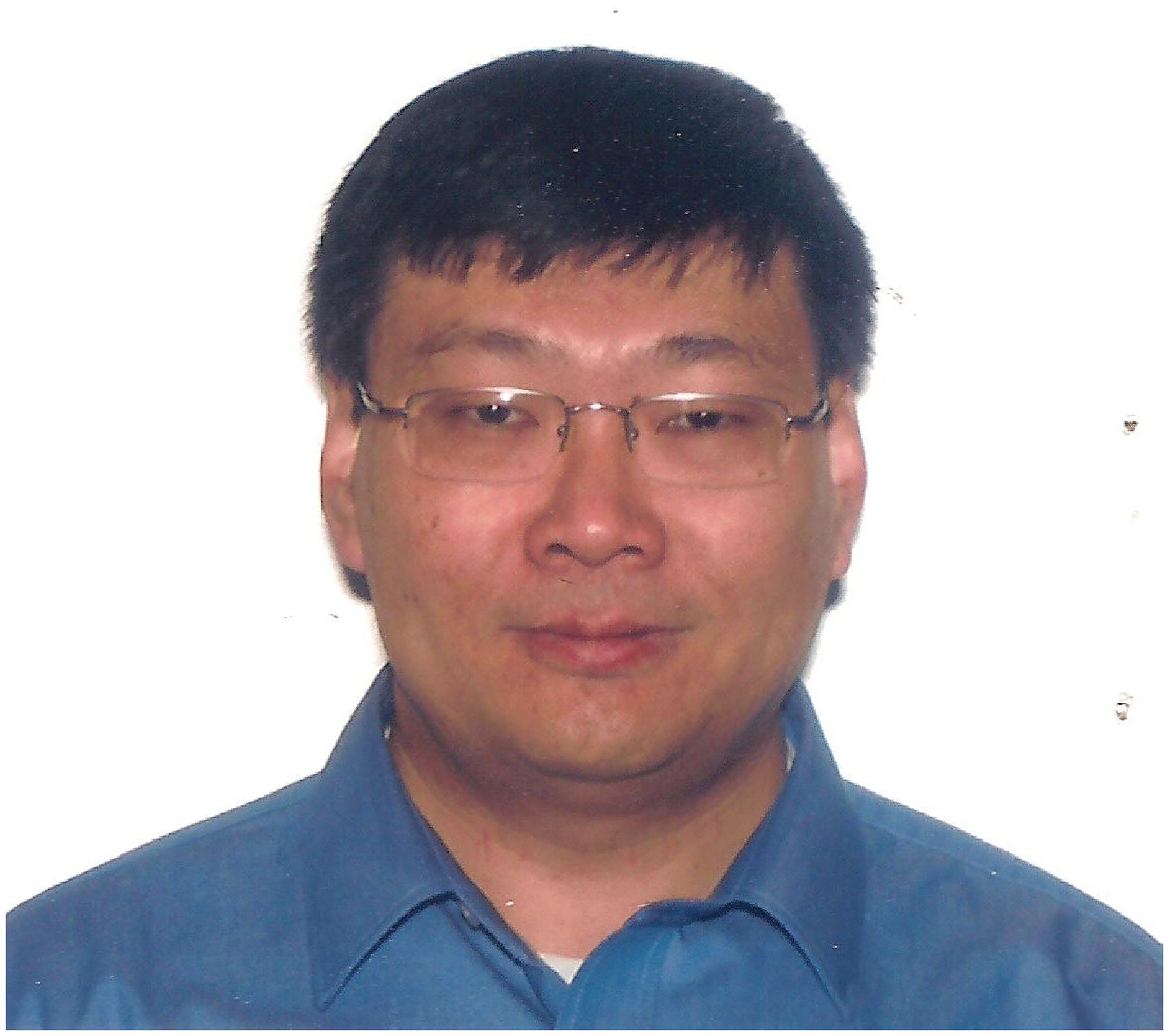}}]{Hang Liu}
joined the Catholic University of America in 2013, where he currently is a Professor with the Department of Electrical Engineering and Computer Science.
Prior to joining the Catholic University, he had more than 10 years of research experience in networking industry and worked in senior research and management positions at several companies.
Dr. Liu has published more than 100 papers in leading journals and conferences, and received two best paper awards and one best student paper award.
He is the inventor/co-inventor of over 50 granted US patents.
He has also made many contributions to the IEEE 802 wireless standards and 3GPP standards, and was the editor of the IEEE 802.11aa standard and the rapporteur of a 3GPP work item.
Dr. Liu received his Ph.D. degree in Electrical Engineering from the University of Pennsylvania.
His research interests include wireless communications and networking, millimeter wave communications, dynamic spectrum management, mobile computing, Internet of Things, future Internet architecture and protocols, mobile content distribution, video streaming, and network security.
He is an IEEE senior member.
\end{IEEEbiography}

\begin{IEEEbiography}[{\includegraphics[width=1in, height=1.25in, clip]{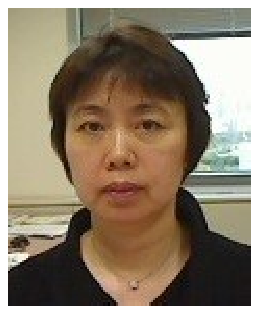}}]{Yusheng Ji}
received her B.E., M.E., and D.E. degrees in electrical engineering from the University of Tokyo.
She joined the National Center for Science Information Systems, Japan (NACSIS) in 1990.
Currently, she is a Professor at the National Institute of Informatics, Japan (NII), and SOKENDAI (the Graduate University for Advanced Studies).
Her research interests include network architecture, mobile computing, and network resource management.
She is/has been an Editor of IEEE Transactions on Vehicular Technology, an Associate Editor of IEICE Transactions and IPSJ Journal, a Guest Editor-in-Chief, a Guest Editor, and a Guest Associate Editor of Special Issues of IEICE Transactions and IPSJ Journal, a Symposium Co-chair of IEEE GLOBECOM 2012 and 2014, a Track Chair of IEEE VTC Fall 2016 and 2017, a General Co-Chair of ICT-DM 2018, and a TPC member of IEEE INFOCOM, ICC, GLOBECOM, WCNC, VTC etc.
She is an IEEE senior member.
\end{IEEEbiography}

\begin{IEEEbiography}[{\includegraphics[width=1in, height=1.25in, clip]{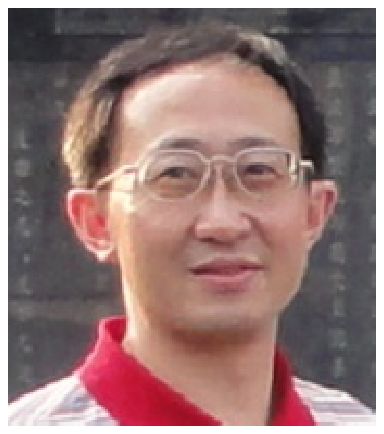}}]{Honggang Zhang}
is a Full Professor with the College of Information Science and Electronic Engineering, Zhejiang University, Hangzhou, China.
He is an Honorary Visiting Professor at the University of York, York, UK.
He was the International Chair Professor of Excellence for Universit\'{e} Europ\'{e}enne de Bretagne (UEB) and Sup\'{e}lec, France.
He is currently active in the research on green communications and was the leading Guest Editor of the IEEE Communications Magazine special issues on ``Green Communications''.
He served as the Series Editor of the IEEE Communications Magazine for the Green Communications and Computing Networks Series from 2015 to 2018 and the Chair of the Technical Committee on Cognitive Networks of the IEEE Communications Society from 2011 to 2012.
He was the co-author and an editor of two books with the titles of \emph{Cognitive Communications Distributed Artificial Intelligence (DAI), Regulatory Policy and Economics, Implementation} (John Wiley \& Sons) and \emph{Green Communications: Theoretical Fundamentals, Algorithms and Applications} (CRC Press), respectively.
He is an Associate Editor-in-Chief of China Communications.
He is an IEEE senior member.
\end{IEEEbiography}


\begin{thebibliography}{29}

\bibitem{Cisc18}
``Cisco visual networking index: Forcast and trends, 2017--2022,'' White Paper, Cisco, Nov. 2018.

\bibitem{Andr12}
J. G. Andrews, H. Claussen, M. Dohler, S. Rangan, and M. C. Reed, ``Femtocells: Past, present, and future,'' \emph{IEEE J. Sel. Areas Commun.}, vol. 30, no. 3, pp. 497--508, Apr. 2012.

\bibitem{Abba18}
N. Abbas, Y. Zhang, A. Taherkordi, and T. Skeie, ``Mobile edge computing: A survey,'' \emph{IEEE Internet Things J.}, vol. 5, no. 1, pp. 450--465, Feb. 2018.

\bibitem{Mao17}
Y. Mao, C. You, J. Zhang, K. Huang and K. B. Letaief, ``A Survey on mobile edge computing: The communication perspective,'' \emph{IEEE Commun. Surveys Tuts.}, vol. 19, no. 4, pp. 2322--2358, Q4 2017.

\bibitem{Saty17}
M. Satyanarayanan, ``The emergence of edge computing,'' \emph{IEEE Comput.}, vol. 50, no. 1, pp. 30--39, Jan. 2017.

\bibitem{Wu19}
Y. Wu, L. P. Qian, K. Ni, C. Zhang, and X. Shen, ``Delay-minimization nonorthogonal multiple access enabled multi-user mobile edge computation offloading,'' \emph{IEEE J. Sel. Topics Signal Process.}, vol. 13, no. 3, pp. 392--407, Jun. 2019.

\bibitem{Zhen19}
J. Zheng, Y. Cai, Y. Wu, and X. Shen, ``Dynamic computation offloading for mobile cloud computing: A stochastic game-theoretic approach,'' \emph{IEEE Trans. Mobile Comput.}, vol. 18, no. 4, pp. 771--786, 1 Apr. 2019.

\bibitem{Zhao18}
P. Zhao, H. Tian, S. Fan, and A. Paulraj, ``Information prediction and dynamic programming based RAN slicing for mobile edge computing,'' \emph{IEEE Wireless Commun. Lett.}, vol. 7, no. 4, pp. 614--617, Aug. 2018.

\bibitem{Zhou17}
Y. Zhou, F. R. Yu, J. Chen, and Y. Kuo, ``Resource allocation for information-centric virtualized heterogeneous networks with in-network caching and mobile edge computing,'' \emph{IEEE Trans. Veh. Technol.}, vol. 66, no. 12, pp. 11339--11351, Dec. 2017.

\bibitem{Xia15}
W. Xia, Y. Wen, C. H. Foh, D. Niyato, and H. Xie, ``A Survey on Software-Defined Networking,'' \emph{IEEE Commun. Surveys Tuts.}, vol. 17, no. 1, pp. 27--51, Q1 2015.

\bibitem{Afol18}
I. Afolabi, T. Taleb, K. Samdanis, A. Ksentini, and H. Flinck, ``Network slicing and softwarization: A survey on principles, enabling technologies, and solutions,'' \emph{IEEE Commun. Surveys Tuts.}, vol. 20, no. 3, pp. 2429--2453, Q3 2018.

\bibitem{Gudi13}
A. Gudipati, D. Perry, L. E. Li, and S. Katti, ``SoftRAN: Software defined radio access network,'' in \emph{ACM SIGCOMM HotSDN Workshop}, Hong Kong, China, Aug. 2013.

\bibitem{Chen14}
T. Chen, H. Zhang, X. Chen, and O. Tirkkonen, ``SoftMobile: Control evolution for future heterogeneous mobile networks,'' \emph{IEEE Wireless Commun.}, vol. 21, no. 6, pp. 70--78, Dec. 2014.

\bibitem{Chen15_1}
T. Chen, M. Matinmikko, X. Chen, X. Zhou, and P. Ahokangas, ``Software defined mobile networks: Concept, survey and research directions,'' \emph{IEEE Commun. Mag.}, vol. 53, no. 11, pp. 126--133, Nov. 2015.

\bibitem{Lian15}
C. Liang and F. R. Yu, ``Wireless network virtualization: A survey, some research issues and challenges,'' \emph{IEEE Commun. Surveys Tuts.}, vol. 17, no. 1, pp. 358--380, Q1 2015.

\bibitem{3gpp14}
``Study on radio access network (RAN) sharing enhancements,'' Rel. 13, 3GPP TR 22.852, Sep. 2014.

\bibitem{3gpp18}
``Telecommunication management; network sharing; concepts and requirements,'' Rel. 15, 3GPP TS 32.130, Jun. 2018.

\bibitem{Samd16}
K. Samdanis, X. Costa-Perez, and V. Sciancalepore, ``From network sharing to multi-tenancy: The 5G network slice broker,'' \emph{IEEE Commun. Mag.}, vol. 54, no. 7, pp. 32--39, Jul. 2016.

\bibitem{Fris08}
T. Frisanco, P. Tafertshofer, P. Lurin, and R. Ang, ``Infrastructure sharing and shared operations for mobile network operators: From a deployment and operations view,'' in \emph{IEEE NOMS}, Salvador, Bahia, Brazil, Apr. 2008.

\bibitem{goog18}
Google, ``Project Fi,'' {https://fi.google.com} [Date Accessed: 12 Dec. 2018].

\bibitem{Ordo17}
J. Ordonez-Lucena, P. Ameigeiras, D. Lopez, J. J. Ramos-Munoz, J. Lorca, and J. Folgueira, ``Network slicing for 5G with SDN/NFV: Concepts, architectures, and challenges,'' \emph{IEEE Commun. Mag.}, vol. 55, no. 5, pp. 80--87, May 2017.

\bibitem{Sall17}
O. Sallent, J. P\'{e}rez-Romero, R. Ferr\'{u}s, and R. Agusti, ``On radio access network slicing from a radio resource management perspective,'' \emph{IEEE Wireless Commun.}, vol. 24, no. 5, pp. 166--174, Oct. 2017.

\bibitem{Shah17}
H. Shah-Mansouri, V. W. S. Wong, and R. Schober, ``Joint optimal pricing and task scheduling in mobile cloud computing systems,'' \emph{IEEE Trans. Wireless Commun.}, vol. 16, no. 8, pp. 5218--5232, Aug. 2017.

\bibitem{ITU18}
``Framework of the IMT-2020 network,'' Rec. ITU-T Y.3102, May 2018.

\bibitem{Zhou16}
X. Zhou, R. Li, T. Chen, and H. Zhang, ``Network slicing as a service: Enabling enterprises' own software-defined cellular networks,'' \emph{IEEE Commun. Mag.}, vol. 54, no. 7, pp. 146--153, Jul. 2016.

\bibitem{Ji07}
Z. Ji and K. J. R. Liu, ``Dynamic spectrum sharing: A game theoretical overview,'' \emph{IEEE Commun. Mag.}, vol. 45, no. 5, pp. 88--94, May 2007.

\bibitem{Hass16}
H. van Hasselt, A. Guez, and D. Silver, ``Deep reinforcement learning with double Q-learning,'' in \emph{Proc. AAAI}, Phoenix, AZ, Feb. 2016.

\bibitem{Abad16}
M. Abadi, P. Barham, J. Chen, Z. Chen, A. Davis, J. Dean, M. Devin, S. Ghemawat, G. Irving, M. Isard, M. Kudlur, J. Levenberg, R. Monga, S. Moore, D. G. Murray, B. Steiner, P. Tucker, V. Vasudevan, P. Warden, M. Wicke, Y. Yu, and X. Zheng, ``Tensorflow: A system for large-scale machine learning,'' in \emph{Proc. OSDI}, Savannah, GA, Nov. 2016.

\bibitem{Han11}
Z. Han, D. Niyato, W. Saad, T. Ba\c{s}ar, and A. Hj{\o}rungnes, \emph{Game Theory in Wireless and Communication Networks: Theory, Models, and Applications}.
Cambridge, UK: Cambridge University Press, 2011.

\bibitem{Han15}
Y. Gu, W. Saad, M. Bennis, M. Debbah, and Z. Han, ``Matching theory for future wireless networks: Fundamentals and applications,'' \emph{IEEE Commun. Mag.}, vol. 53, no. 5, pp. 52--59, May 2015.

\bibitem{Dats18}
E. Datsika, A. Antonopoulos, D. Yuan, and C. Verikoukis, ``Matching theory for over-the-top service provision in 5G networks,'' \emph{IEEE Trans. Wireless Commun.}, vol. 17, no. 8, pp. 5452--5464, Aug. 2018.

\bibitem{Xiao18}
Y. Xiao, M. Hirzallah, and M. Krunz, ``Distributed resource allocation for network slicing over licensed and unlicensed bands,'' \emph{IEEE J. Sel. Areas Commun.}, vol. 36, no. 10, pp. 2260--2274, Oct. 2018.

\bibitem{Caba19}
P. Caballero, A. Banchs, G. de Veciana, and X. Costa-P\'{e}rez, ``Network slicing games: Enabling customization in multi-tenant networks,'' \emph{IEEE/ACM Trans. Netw.}, Early Access Article, Feb. 2019.

\bibitem{DOro18}
S. D'Oro, F. Restuccia, T. Melodia, and S. Palazzo, ``Low-complexity distributed radio access network slicing: Algorithms and experimental results,'' \emph{IEEE/ACM Trans. Netw.}, vol. 26, no. 6, pp. 2815--2828, Dec. 2018.

\bibitem{Sun19}
Y. Sun, M. Peng, S. Mao, and S. Yan, ``Hierarchical radio resource allocation for network slicing in fog radio access networks,'' \emph{IEEE Trans. Veh. Technol.}, Early Access Article, Jan. 2019.

\bibitem{Xiao1802}
Y. Xiao and M. Krunz, ``Dynamic network slicing for scalable fog computing systems with energy harvesting,'' \emph{IEEE J. Sel. Areas Commun.}, vol. 36, no. 12, pp. 2640--2654, Dec. 2018.

\bibitem{Fu13}
F. Fu and U. C. Kozat, ``Stochastic game for wireless network virtualization,'' \emph{IEEE/ACM Trans. Netw.}, vol. 21, no. 1, pp. 84--97, Feb. 2013.

\bibitem{Jian17}
C. Jiang, H. Zhang, Y. Ren, Z. Han, K. Chen, and L. Hanzo, ``Machine learning paradigms for next-generation wireless networks,'' \emph{IEEE Wireless Commun.}, vol. 24, no. 2, pp. 98--105, Apr. 2017.

\bibitem{ML5G}
``Focus Group on Machine Learning for Future Networks including 5G,'' \url{https://www.itu.int/en/ITU-T/focusgroups/ml5g/Pages/default.aspx}.

\bibitem{ETSI}
``Experiential Network Intelligence (ENI),'' \url{https://www.etsi.org/technologies/experiential-networked-intelligence}.

\bibitem{ISO}
``Artificial Intelligence,'' ISO/IEC JTC 1/SC 42, \url{https://www.iso.org/committee/6794475.html}.

\bibitem{TM}
``Artificial Intelligence makes Smart BPM Smarter,'' TM Forum Catalyst Project, \url{https://www.tmforum.org/catalysts/smart-bpm/}.

\bibitem{Ho14}
C. Ho, D. Yuan, and S. Sun, ``Data offloading in load coupled networks: A utility maximization framework," \emph{IEEE Trans. Wireless Commun.}, vol. 13, no. 4, pp. 1921--1931, Apr. 2014.

\bibitem{Chen15}
X. Chen, J. Wu, Y. Cai, H. Zhang, and T. Chen, ``Energy-efficiency oriented traffic offloading in wireless networks: A brief survey and a learning approach for heterogeneous cellular networks,'' \emph{IEEE J. Sel. Areas Commun.}, vol. 33, no. 4, pp. 627--640, Apr. 2015.

\bibitem{Nich08}
A. J. Nicholson and B. D. Noble, ``BreadCrumbs: Forecasting mobile connectivity,'' in \emph{Proc. ACM MobiCom}, San Francisco, CA, Sep. 2008.

\bibitem{Cheu15}
M. H. Cheung and J. Huang, ``DAWN: Delay-aware Wi-Fi offloading and network selection,'' \emph{IEEE J. Sel. Areas Commun.}, vol. 33, no. 6, pp. 1214--1223, Jun. 2015.

\bibitem{Jia09}
J. Jia, Q. Zhang, Q. Zhang, and M. Liu, ``Revenue generation for truthful spectrum auction in dynamic spectrum access,'' in \emph{Proc. ACM MobiHoc}, New Orleans, LA, May 2009.

\bibitem{He17}
X. He, J. Liu, R. Jin, and H. Dai, ``Privacy-aware offloading in mobile-edge computing,'' in \emph{Proc. IEEE GLOBECOM}, Singapore, Dec. 2017.

\bibitem{Berr02}
R. A. Berry and R. G. Gallager, ``Communication over fading channels with delay constraints,'' \emph{IEEE Trans. Inf. Theory}, vol. 48, no. 5, pp. 1135--1149, May 2002.

\bibitem{Burd96}
T. D. Burd and R. W. Brodersen, ``Processor design for portable systems,'' \emph{J. VLSI Signal Process. Syst.}, vol. 13, no. 2–3, pp. 203--221, Aug. 1996.

\bibitem{Fink64}
A. M. Fink, ``Equilibrium in a stochastic $n$-person game,'' \emph{J. Sci. Hiroshima Univ. Ser. A-I}, vol. 28, pp. 89--93, 1964.

\bibitem{Chen1801}
X. Chen, Z. Han, H. Zhang, G. Xue, Y. Xiao, and M. Bennis, ``Wireless resource scheduling in virtualized radio access networks using stochastic learning,'' \emph{IEEE Trans. Mobile Comput.}, vol. 17, no. 4, pp. 961--974, Apr. 2018.

\bibitem{Kroe16}
C. Kroer and T. Sandholm, ``Imperfect-recall abstractions with bounds in games,'' in \emph{Proc. ACM EC}, Maastricht, the Netherlands, Jul. 2016.

\bibitem{Abel16}
D. Abel, D. Hershkowitz, and M. Littman, ``Near optimal behavior via approximate state abstraction,'' in \emph{Proc. ICML}, New York, NY, Jun. 2016.

\bibitem{Fu09}
F. Fu and M. van der Schaar, ``Learning to compete for resources in wireless stochastic games,'' \emph{IEEE Trans. Veh. Technol.}, vol. 58, no. 4, pp. 1904--1919, May 2009.

\bibitem{Tsit96}
J. N. Tsitsiklis and B. van Roy, ``Feature-based methods for large scale dynamic programming,'' \emph{Mach. Learn.}, vol. 22, no. 1-3, pp. 59--94, Jan. 1996.

\bibitem{Rich98}
R. S. Sutton and A. G. Barto, \emph{Reinforcement Learning: An Introduction}. Cambridge, MA: MIT Press, 1998.

\bibitem{Loev77}
M. Lo\`{e}ve, \emph{Probability Theory I}. Berlin, Germany:  Springer-Verlag, 1977.

\bibitem{Watk12}
C. J. C. H. Watkins and P. Dayan, ``Q-learning,'' \emph{Mach. Learn.}, vol. 8, no. 3--4, pp. 279--292, May 1992.

\bibitem{Appl17}
Apple, ``The future is here: iPhone X,'' \url{https://www.apple.com/newsroom/2017/09/the-future-is-here-iphone-x/} [Date Accessed: 16 Jul. 2018].

\bibitem{Mnih15}
V. Mnih, K. Kavukcuoglu, D. Silver, A. A. Rusu, J. Veness, M. G. Bellemare, A. Graves, M. Riedmiller, A. K. Fidjeland, G. Ostrovski, S. Petersen, C. Beattie, A. Sadik, I. Antonoglou, H. King, D. Kumaran, D. Wierstra, S. Legg, and D. Hassabis, ``Human-level control through deep reinforcement learning,'' \emph{Nature}, vol. 518, no. 7540, pp. 529--533, Feb. 2015.

\bibitem{Lin92}
L.-J. Lin, ``Reinforcement learning for robots using neural networks,'' Carnegie Mellon University, 1992.

\bibitem{Siom12}
I. Siomina and D. Yuan, ``Analysis of Cell Load Coupling for LTE Network Planning and Optimization,'' \emph{IEEE Trans. Wireless Commun.}, vol. 11, no. 6, pp. 2287--2297, Jun. 2012.

\bibitem{Mao16}
Y. Mao, J. Zhang, and K. B. Letaief, ``Dynamic computation offloading for mobile-edge computing with energy harvesting devices,'' \emph{IEEE J. Sel. Areas Commun.}, vol. 34, no. 12, pp. 3590--3605, Dec. 2016.

\bibitem{CCLP18}
C. L. P. Chen and Z. Liu, ``Broad learning system: An effective and efficient incremental learning system without the need for deep architecture,'' \emph{IEEE Trans. Neural Netw. Learn. Syst.}, vol. 29, no. 1, pp. 10--24, Jan. 2018.

\bibitem{Chen1802}
X. Chen, H. Zhang, C. Wu, S. Mao, Y. Ji, and M. Bennis, ``Performance optimization in mobile-edge computing via deep reinforcement learning,'' in \emph{Proc. IEEE VTC}, Chicago, IL, Aug. 2018.

\bibitem{Jarr09}
K. Jarrett, K. Kavukcuoglu, M. Ranzato, and Y. LeCun, ``What is the best multi-stage architecture for object recognition?'' in \emph{Proc. IEEE ICCV}, Kyoto, Japan, Sep.--Oct. 2009.

\bibitem{King15}
D. P. Kingma and J. Ba, ``Adam: A Method for Stochastic Optimization,'' in \emph{Proc. ICLR}, San Diego, CA, May 2015.

\end{thebibliography}
\end{document}